\documentclass{SciPost}

\hypersetup{
    colorlinks,
    linkcolor={red!50!black},
    citecolor={blue!50!black},
    urlcolor={blue!80!black}
}

\usepackage[bitstream-charter]{mathdesign}
\DeclareSymbolFont{usualmathcal}{OMS}{cmsy}{m}{n}
\DeclareSymbolFontAlphabet{\mathcal}{usualmathcal}

\fancypagestyle{SPstyle}{
\fancyhf{}
\lhead{\colorbox{scipostblue}{\bf \color{white} ~SciPost Physics }}
\rhead{{\bf \color{scipostdeepblue} ~Submission }}

\fancyfoot[C]{\textbf{\thepage}}
}

\usepackage{mathtools}
\usepackage{adjustbox}
\usepackage{enumitem}
\usepackage{tikz}

\newcommand{\Int}{\mathbb{Z}}

\DeclareMathOperator{\Tr}{tr}
\DeclareMathOperator{\Det}{det}

\DeclarePairedDelimiterX{\innerp}[2]{\langle}{\rangle}{#1, #2}
\DeclarePairedDelimiter{\bra}{\langle}{\rvert}
\DeclarePairedDelimiter{\ket}{\lvert}{\rangle}
\DeclarePairedDelimiterX{\braket}[2]{\langle}{\rangle}{#1 \delimsize\vert #2}
\DeclarePairedDelimiterX{\ketbra}[2]{\lvert}{\rvert}{#1 \delimsize\rangle\!\delimsize\langle #2}
\DeclarePairedDelimiterX{\proj}[1]{\lvert}{\rvert}{#1 \delimsize\rangle\!\delimsize\langle #1}

\tikzset{
  every picture/.style={baseline,thick},
  base/.style = {draw, rectangle, rounded corners, fill=black!5}
}

\begin{document}

\pagestyle{SPstyle}

\begin{center}{\Large \textbf{\color{scipostdeepblue}{
Two-parameter families of matrix product operator\\
integrals of motion in Heisenberg spin chains
}}}\end{center}

\begin{center}\textbf{
Vsevolod I. Yashin\textsuperscript{1,2$\star$}
}\end{center}

\begin{center}
{\bf 1} Steklov Mathematical Institute of Russian Academy of Sciences, Moscow 119991, Russia
\\
{\bf 2} Russian Quantum Center, Skolkovo, Moscow 143025, Russia
\\[\baselineskip]
$\star$ \href{mailto:email1}{\small yashin.vi@mi-ras.ru}
\end{center}

\section*{\color{scipostdeepblue}{Abstract}}
\textbf{\boldmath{%
  Recently, Fendley et al. (2025) revealed a new simple way to demonstrate the integrability of XYZ Heisenberg model by constructing a one-parameter family of integrals of motion in the matrix product operator (MPO) form with bond dimension $4$. In this work, I report on the discovery of two-parameter families of MPOs that commute with Heisenberg spin chain Hamiltonian in case of various anisotropies (XXX, XXZ, XX, XY and XYZ). These solutions are connected by taking appropriate limits. For all cases except XYZ, I also write down Floquet charges of two-step Floquet protocols corresponding to the Trotterization. I describe a symbolic algebra approach for finding such integrals of motion and speculate about possible generalizations and applications.
}}

\vspace{\baselineskip}



\vspace{10pt}
\noindent\rule{\textwidth}{1pt}
\tableofcontents
\noindent\rule{\textwidth}{1pt}
\vspace{10pt}


\section{Introduction}
\label{sec:introduction}

One-dimensional Heisenberg spin chain models \cite{Heisenberg_1928} are very popular exactly solvable models for studying critical points and phase transitions of magnetic systems in quantum many-body physics. There is a long and impactful history of mathematical and physical breakthroughs connected with Heisenberg models. Early developments include Hans Bethe's solution to isotropic Heisenberg model using his famous ansatz \cite{Bethe_1931} and its generalization to XXZ anisotropic case \cite{Orbach_1958,Yang_1966_1,Yang_1966_2,Yang_1966_3}. Later, Sutherland found the connection of XYZ Heisenberg model with classical eight-vertex model \cite{Sutherland_1970}, and Baxter managed to solve these models \cite{Baxter_1973_1,Baxter_1973_2,Baxter_1973_3}. The methods developed in these works evolved to become the algebraic Bethe ansatz approach \cite{Takhtadzhyan_1979,Slavnov_2019} and were applied to wide range of integrable models \cite{Baxter_2007}.

Bethe ansatz methods do not directly lead to concrete description of the local conserved quantities of the model. The question of explicitly constructing local charges (higher Hamiltonians) was solved for isotropic XXX Heisenberg spin chain by Grabowski and Mathieu \cite{Grabowski_1994,Grabowski_1995}. They used boost operator and combinatorial identitites to construct explicit form of local charges in terms of recursive relations. In \cite{Yamada_2023}, Yamada and Fukai encoded these conserved quantities using matrix product operator (MPO) \cite{Verstraete_2004,Zwolak_2004,Pirvu_2010} form. Anisotropic XXZ model allows for description in terms of Temperley-Lieb algebra, Nienhuis and Huijgen gave a closed-form description to XXZ local charges in terms of Temperley-Lieb algebra elements \cite{Nienhuis_2021}. Nozawa and Fukai worked out the complete description via recursion relations to local charges of XYZ model \cite{Nozawa_2020}.

One modern direction in integrable models concerns studying spin chains with periodic driving. Some systems with periodically time-dependent Hamiltonians exhibit non-chaotic behaviour and come with quantities that are conserved up to a time period (Floquet charges). For example, such systems naturally appear after Trotterization of non-driven integrable models. Studying Floquet integrable models was popularized by Gritsev and Polkovnikov \cite{Gritsev_2017}, but some of the corresponding constructions appeared much earlier in light-cone discretizations of field theories \cite{Destri_1987,Faddeev_1994}. Ljubotina, Zadnik and Prosen studied two-step brick-wall model given by Trotterized XXZ spin chain and found a family of quasi-local charges for it \cite{Ljubotina_2019}. The two-step Trotterized XXZ model with anisotropy parameter $\Delta=i$ was also studied in \cite{Yashin_2023}, where the Temperley-Lieb algebra was used to give closed-form expressions to Floquet local integrals of motion. The Floquet integrability of quantum Potts models was shown in \cite{Lotkov_2022}. Miao, Gritsev and Kurlov proved the Floquet integrability of wide range of models satisfying Yang-Baxter relations \cite{Miao_2024}. Nowadays, the Floquet integrability is studied for great variety of quantum circuit geometries \cite{Vanicat_2017,Miao_2023,Zadnik_2024,Richelli_2024,Paletta_2025,Znidaric_2025} and is related to the phenomenon of dynamical freezing \cite{Haldar_2021,Lu_2025,Mukherjee_2026}.

Recently, Paul Fendley, Sascha Gehrmann, Eric Vernier and Frank Verstraete found a simplified way to show the integrability of the XYZ Heisenberg spin chain \cite{Fendley_2025}. They proposed to study integrals of motion in the MPO form and formulated a simple sufficient condition on local tensors that makes MPO commute with the Hamiltonian. They gave an explicit MPO solution with bond dimension $4$ dependent on one parameter and satisfying these conditions (generalizing the MPO family of charges for XXX model found by Katsura \cite{Katsura_2015}). Expanding the MPO charge as a series over this parameter, they obtained even local charges of the XYZ model. Their work also contains results about the relation between MPO charges and Bethe ansatz transfer matrices, and generalizations to various boundary conditions and defects. In general, it appears that the scope of the method should be broad. Even more recently, Fukai and Yamada \cite{Fukai_2026} constructed a one-parameter family of MPO integrals of motion that contains the information about all local charges of the XYZ model. This solution helped them to find a better explanation to the combinatorial structure of the local charges. Their MPO representation uses dual number $\varepsilon$ such that $\varepsilon^2=0$, which can be rewritten as the usual MPO with bond dimension higher than $4$ by presenting the dual number as $2\times 2$ matrix.

In this work, I exploit the methods of \cite{Fendley_2025} and some symbolic algebra to find two-parameter families of MPO integrals of motions with bond dimension $4$ applicable to XXX, XXZ, XX, XY and XYZ Heisenberg spin chains. Expanding these solutions into series near some zero point, one can obtain local charges of the spin chains, and I conjecture that the solutions might contain information about all local integrals of motion. At least for XXX and XXZ models, the parameters can be chosen to have rather convenient geometry (they naturally depend on points of a sphere). Also, I show that all models except XYZ are stable under Trotterization, by explicitly writing down two-parameter families of MPO Floquet charges for two-step brick-wall Floquet protocols. The found family of MPO integrals of motion for XYZ Heisenberg spin chain depends on points of projective plane, and MPO families for other models can be obtained as limiting cases. Unlike Baxter's solution of eight-vertex model, the found MPO conserved quantities do not require using the theory of special functions. The work was done independently of \cite{Fukai_2026}.

This paper is organized as follows. After giving concise preliminaries about quantum Heisenberg spin chains and their their MPO integrals of motion (Section~\ref{sec:preliminaries}), I state the results, which consist in writing down two-parameter families of integrals of motion for XXX, XXZ, XX, XY and XYZ Heisenberg models (Section~\ref{sec:results}). Then I disclose two basic ideas that helped me during the search (Section~\ref{sec:methods}), and finally I conclude and elaborate on possible future research and applications (Section~\ref{sec:discussion}).

\section{Preliminaries}
\label{sec:preliminaries}

Let me start by giving minimal preliminaries necessary for formulating the results.

\subsection{Hamiltonians}
\label{subsec:hamiltonians}

Suppose there are $L$ spin-$\frac{1}{2}$ sites, each site labeled by coordinate $j \in \{1,\dots,L\}$. I denote Pauli operators acting on $j$-th site as $\{\mathbf{X}_j,\mathbf{Y}_j,\mathbf{Z}_j\}$ so that $\mathbf{X}_j^2 =\,\mathbf{Y}_j^2 =\,\mathbf{Z}_j^2 = 1,\;\,\mathbf{X}_j\,\mathbf{Y}_j\,\mathbf{Z}_j = i$ for all $j$ (here $1$ is an identity operator and $i$ is an imaginary unit), and the operators on different sites commute.

I will condsider one-dimensional Heisenberg models with periodic boundary conditions. The Hamiltonians of such models in isotropic XXX and anisotropic XXZ, XX, XY and XYZ cases are defined as
\begin{equation}
\label{eq:hamiltonians}
\begin{aligned}
  H^{XXX}&=\sum_{j=1}^L H_{j,j+1}^{XXX}, \quad & H_{j,j+1}^{XXX} &= \frac{1}{2}\left[ \mathbf{X}_j \mathbf{X}_{j+1} + \mathbf{Y}_j \mathbf{Y}_{j+1} + \mathbf{Z}_j \mathbf{Z}_{j+1} \right], \\
  H^{XXZ}&=\sum_{j=1}^L H_{j,j+1}^{XXZ}, \quad & H_{j,j+1}^{XXZ} &= \frac{1}{2}\left[ \mathbf{X}_j \mathbf{X}_{j+1} + \mathbf{Y}_j \mathbf{Y}_{j+1} + \Delta\,\mathbf{Z}_j \mathbf{Z}_{j+1} \right], \\
  H^{XX}&=\sum_{j=1}^L H_{j,j+1}^{XX}, \quad & H_{j,j+1}^{XX} &= \frac{1}{2}\left[ \mathbf{X}_j \mathbf{X}_{j+1} + \mathbf{Y}_j \mathbf{Y}_{j+1} \right], \\
  H^{XY}&=\sum_{j=1}^L H_{j,j+1}^{XY}, \quad & H_{j,j+1}^{XY} &= \frac{1}{2}\left[ J_x\,\mathbf{X}_j \mathbf{X}_{j+1} + J_y\,\mathbf{Y}_j \mathbf{Y}_{j+1} \right],\\
  H^{XYZ}&=\sum_{j=1}^L H_{j,j+1}^{XYZ}, \quad & H_{j,j+1}^{XYZ} &= \frac{1}{2}\left[ J_x\,\mathbf{X}_j \mathbf{X}_{j+1} + J_y\,\mathbf{Y}_j \mathbf{Y}_{j+1} + J_z\,\mathbf{Z}_j \mathbf{Z}_{j+1} \right], \\
\end{aligned}
\end{equation}
where $\Delta$ and $J_x,J_y,J_z$ are coupling constants. These models transition one into another when taking different parameter limits. For simplicity, I study translationally invariant Heisenberg models with periodic boundary conditions, which means $L+1\equiv 1$.

\subsection{Matrix product operators}
\label{subsec:matrix_product_operators}

Given a Hamiltonian $H$, any operator $\mathcal{M}$ commuting with Hamiltonian $[H,\mathcal{M}]=0$ remains constant during the evolution and is called an \emph{integral of motion} or a \emph{charge}. In this work, I search for integrals of motion given in a \emph{matrix product operator (MPO)} \cite{Verstraete_2004,Zwolak_2004,Pirvu_2010} form
\begin{equation}
\label{eq:MPO_operator}
  \mathcal{M} = \Tr[\mathcal{A}_1 \mathcal{A}_2 \cdots \mathcal{A}_L],
\end{equation}
where each $\mathcal{A}_j$ is a $4\times 4$-matrix with given by operators acting on site $j$, each entry is defined by two indices from $\{0,x,y,z\}$:
\begin{equation}
\label{eq:arbitrary_operator_matrix}
  \mathcal{A}_j =
  \begin{bmatrix}
    (\mathcal{A}_j)_{0,0} & (\mathcal{A}_j)_{0,x} & (\mathcal{A}_j)_{0,y} & (\mathcal{A}_j)_{0,z} \\
    (\mathcal{A}_j)_{x,0} & (\mathcal{A}_j)_{x,x} & (\mathcal{A}_j)_{x,y} & (\mathcal{A}_j)_{x,z} \\
    (\mathcal{A}_j)_{y,0} & (\mathcal{A}_j)_{y,x} & (\mathcal{A}_j)_{y,y} & (\mathcal{A}_j)_{y,z} \\
    (\mathcal{A}_j)_{z,0} & (\mathcal{A}_j)_{z,x} & (\mathcal{A}_j)_{z,y} & (\mathcal{A}_j)_{z,z}
  \end{bmatrix}
  .
\end{equation}
Row and column dimensions of such matrices are called \emph{bond dimensions}, in the described case they equal $4$. Multiplication of two operator matrices is defined as the usual matrix multiplication and the trace $\Tr$ is a sum of diagonal entries. Note that operator matrices $\mathcal{A}_j$ can be multiplied by scalars, such renormalizations are physically irrelevant. Because I consider translationally invariant models, the operator matrix $\mathcal{A}_j$ remains constant over all sites $j$. Later in text I will leave lower index and write $\mathcal{A}$ for visual clarity.

It is customary and illustrative to depict operator matrices $\mathcal{A}$ as \emph{tensor diagrams} \cite{Orus_2014,Biamonte_2020,Cuiper_2026}
\begin{equation}
  \bra{n}(\mathcal{A}_j)_{\alpha,\beta}\ket{m} =
  \adjustbox{valign=m}{
  \begin{tikzpicture}
    \draw (0,0) -- (0,0.8) node[above] {$n$};
    \draw (0,0) -- (0,-0.8) node[below] {$m$};
    \draw (0,0) -- (-0.8,0) node[left] {$\alpha$};
    \draw (0,0) -- (0.8,0) node[right] {$\beta$};
    \node[base,fill=blue!10] at (0,0) {$\mathcal{A}_j$};
  \end{tikzpicture}
  }
  ,
\end{equation}
where  vertical lines represent \emph{physical indices} $n,m$ and horizontal lines represent \emph{virtual indices} $\alpha,\beta$. Connecting lines between diagrams corresponds to contracting the tensor index. The tensor diagram of the matrix product operator $\mathcal{M}$ with periodic boundary conditions reads
\begin{equation}
  \mathcal{M} =
  \,
  \adjustbox{valign=m}{
  \begin{tikzpicture}
    \
    \draw[rounded corners] (0.4,0) -- (8,0) node[fill=white] {$\cdots$} -- (9.6,0)
      -- (9.6,-0.5) --  (0.4,-0.5) -- cycle;
    \foreach \x in {1,...,7,9}
      \draw (\x,-0.8) -- (\x,0.8);
    \foreach \x in {1,...,7}
      \node[base,fill=blue!10] at (\x,0) {$\mathcal{A}_\x$};
    \node[base,fill=blue!10] at (9,0) {$\mathcal{A}_L$};
  \end{tikzpicture}
  }.
\end{equation}

\subsection{Main equation}
\label{subsec:main_equation}

Heisenberg model Hamiltonians [Eq.~\eqref{eq:hamiltonians}] consist of nearest-neighbour interactions and each Hamiltonian $H$ is decomposed as $H = \sum_{j} H_{j,j+1}$, where $H_{j,j+1}$ is local term acting of two sites $j,j+1$. Recently, Paul Fendley, Sascha Gehrmann, Eric Vernier and Frank Verstraete proposed to study a simple sufficient condition for MPO integrals of motion \cite{Fendley_2025}: for all sites $j$ with operator matrices $\mathcal{A}_j$ on them, there should exist operator matrices $\mathcal{E}_j$ called \emph{error terms} satisfying
\begin{equation}
\label{eq:main_equation}
  i \;
  \adjustbox{raise=0.25em}{
  \begin{tikzpicture}
    \draw (-0.7,0) -- (1.7,0);
    \draw (0,-0.7) -- (0,1.6); \draw (1,-0.7) -- (1,1.6);
    \draw[base] (-0.3,0.6) rectangle (1.3,1.3) node[midway] {$H_{j,j+1}$};
    \node[base,fill=blue!10] at (0,0) {$\mathcal{A}_j$};
    \node[base,fill=blue!10] at (1,0) {$\mathcal{A}_{j\!+\!1}$};
  \end{tikzpicture}
  }
  \;
  -
  i \;
  \adjustbox{raise=0.25em}{
  \begin{tikzpicture}
    \draw (-0.7,0) -- (1.7,0);
    \draw (0,0.7) -- (0,-1.6); \draw (1,0.7) -- (1,-1.6);
    \node[base,fill=blue!10] at (0,0) {$\mathcal{A}_j$};
    \node[base,fill=blue!10] at (1,0) {$\mathcal{A}_{j\!+\!1}$};
    \draw[base] (-0.3,-0.6) rectangle (1.3,-1.3) node[midway] {$H_{j,j+1}$};
  \end{tikzpicture}
  }
  \quad = \quad
  \adjustbox{raise=0.25em}{
  \begin{tikzpicture}
    \draw (-0.7,0) -- (1.7,0);
    \draw (0,-0.7) -- (0,0.7); \draw (1,-0.7) -- (1,0.7);
    \node[base,fill=yellow!10] at (0,0) {$\mathcal{E}_j$} ;
    \node[base,fill=blue!10] at (1,0) {$\mathcal{A}_{j\!+\!1}$};
  \end{tikzpicture}
  }
  -
  \adjustbox{raise=0.25em}{
  \begin{tikzpicture}
    \draw (-0.7,0) -- (1.7,0);
    \draw (0,-0.7) -- (0,0.7); \draw (1,-0.7) -- (1,0.7);
    \node[base,fill=blue!10] at (0,0) {$\mathcal{A}_j$};
    \node [base, fill=yellow!10] at (1,0) {$\mathcal{E}_{j+1}$} ;
  \end{tikzpicture}
  }
  .
  \tag{$\bigstar$}
\end{equation}
By completing the diagrams with local tensors to construct full MPO $\mathcal{M}$ and summing over $j$, error terms cancel, giving $[\mathcal{M},H] = 0$. Note that the error terms are not unique: if $\mathcal{E}_j$ satisfies Eq.~\eqref{eq:main_equation}, then $\mathcal{E}_j + \gamma \mathcal{A}_j$ also satisfies Eq.~\eqref{eq:main_equation} for any value $\gamma$.

So, in order to find MPO integrals of motion, one might search for solutions of Eq.~\eqref{eq:main_equation}. Since I will be working with translationally invariant systems, the elements occuring the equation will be independent of $j$.

\subsection{Trotterized dynamics}
\label{subsec:trotterized_dynamics}

It is sometimes inetesting to study Trotterized spin chain dynamics, which in the usual case is given by two-step Floquet protocol: assuming periodic boundary conditions and that the number of sites $L$ is even, there are two (``even'' and ``odd'') Hamiltonians
\begin{equation}
  H_o = \sum_{j \text{ odd}} H_{j,j+1}, \qquad H_e = \sum_{j \text{ even}} H_{j,j+1},
\end{equation}
that interchangeably act on a spin chain for time periods $\tau$, assembling a brick-wall quantum circuit:
\begin{equation}
\label{eq:brick-wall}
  \adjustbox{valign=m,scale=0.8}{
  \begin{tikzpicture}
    \foreach \x in {1,...,15} {
      \draw (\x,-0.7) node[anchor=north,yshift=0.5em] {$\vdots$} -- (\x,3.7) node[anchor=south] {$\vdots$};
    }
    \foreach \y in {0,2} {
      \foreach \x [evaluate=\x as \xs using int(\x+1)] in {1,3,...,14}
        \draw[base] (\x-0.3,\y-0.35) rectangle (\xs+0.3,\y+0.35) node[midway] {$e^{-i\tau H_{\x,\xs}}$};
      \node[fill=white] at (15,\y) {$\cdots$};
    }
    \foreach \y in {1,3} {
      \node[fill=white] at (1,\y) {$\cdots$};
      \foreach \x [evaluate=\x as \xs using int(\x+1)] in {2,4,...,14}
        \draw[base] (\x-0.3,\y-0.35) rectangle (\xs+0.3,\y+0.35) node[midway] {$e^{-i\tau H_{\x,\xs}}$};
    }
  \end{tikzpicture}
  }
  .
\end{equation}
In the limit $\tau\to 0$ this circuit reproduces dynamics of the model with time-independent Hamiltonian $H$. An observable $\mathcal{M}_\tau$ is called a \emph{Floquet integral of motion} \cite{Gritsev_2017} if it is conserved up to a time period, which means
\begin{equation}
  e^{i\tau H_e}e^{i\tau H_o}\,\mathcal{M}_\tau\,e^{-i\tau H_o}e^{-i\tau H_e} = \mathcal{M}_\tau.
\end{equation}
For two-step Trotterizations of translationally invariant models, it is natural to search Floquet integrals of motion in the shift-by-two-sites form given by local tensors of two types $\mathcal{A}_o$ and $\mathcal{A}_e$:
\begin{equation}
  \mathcal{M}_\tau =
  \,
  \adjustbox{valign=m}{
  \begin{tikzpicture}
    \draw[rounded corners] (0.4,0) -- (8,0) node[fill=white] {$\cdots$} -- (9.6,0)
      -- (9.6,-0.5) --  (0.4,-0.5) -- cycle;
    \foreach \x in {1,...,7,9}
      \draw (\x,-0.8) -- (\x,0.8);
    \foreach \x in {1,3,...,7}
      \node[base,fill=blue!10] at (\x,0) {$\mathcal{A}_o$};
    \foreach \x in {2,4,...,7}
      \node[base,fill=violet!10] at (\x,0) {$\mathcal{A}_e$};
    \node[base,fill=violet!10] at (9,0) {$\mathcal{A}_e$};
  \end{tikzpicture}
  }.
\end{equation}
In the Trotter's limit $\tau\to 0$, it is naturally to expect that the operator matrices $\mathcal{A}_o$ and $\mathcal{A}_e$ should coincide.

The Floquet analogue of Eq.~\eqref{eq:main_equation} states that local evolutions should swap $\mathcal{A}_o$ and $\mathcal{A}_e$:
\begin{equation}
\label{eq:trotterized_main_equation}
  \adjustbox{raise=0.25em}{
  \begin{tikzpicture}
    \draw (-0.7,0) -- (1.7,0);
    \draw (0,-1.6) -- (0,1.6); \draw (1,-1.6) -- (1,1.6);
    \draw[base] (-0.3,0.6) rectangle (1.3,1.3) node[midway] {$e^{i\tau H_{j,j+1}}$};
    \draw[base] (-0.3,-0.6) rectangle (1.3,-1.3) node[midway] {$e^{-i\tau H_{j,j+1}}$};
    \node[base,fill=blue!10] at (0,0) {$\mathcal{A}_o$};
    \node[base,fill=violet!10] at (1,0) {$\mathcal{A}_e$};
  \end{tikzpicture}
  }
  \quad = \quad
  \adjustbox{raise=0.25em}{
  \begin{tikzpicture}
    \draw (-0.7,0) -- (1.7,0);
    \draw (0,-0.7) -- (0,0.7); \draw (1,-0.7) -- (1,0.7);
    \node[base,fill=violet!10] at (0,0) {$\mathcal{A}_e$};
    \node[base,fill=blue!10] at (1,0) {$\mathcal{A}_o$};
  \end{tikzpicture}
  }.
  \tag{$\blacklozenge$}
\end{equation}
Then, the evolution of the MPO consists of swapping $\mathcal{A}_o$ and $\mathcal{A}_e$ after each layer. Some authors reasonably prefer to call this condition the Yang-Baxter equation \cite{Zadnik_2024,Paletta_2025}. When $\tau\to 0$, this equation tends to Eq.~\eqref{eq:main_equation}, where the error terms $\mathcal{E}$ indicate the discrepancy between $\mathcal{A}_e$ and $\mathcal{A}_o$.

So, in order to find Floquet integrals of motion in Trotterized dynamics, one might search for solutions of Eq.~\eqref{eq:trotterized_main_equation}.

\section{Results}
\label{sec:results}

This Section contains results of the work. I will gradually list solutions to Eq.~\eqref{eq:main_equation} for XXX, XXZ, XX, XY and finally XYZ model. Additionally, for all models except XYZ, I solve Eq.~\eqref{eq:trotterized_main_equation}. All these solutions will depend on two-dimensional spaces of parameters (excluding scaling).

One can make sure that the listed solutions are correct by direct substitution. I provide  \textrm{Mathematica} notebooks \cite{Mathematica} with corresponing checks in Supplementary Materials \cite{Supplementary}. I give a comment on what methods I used to find the solutions in Section~\ref{sec:methods}.

\subsection{XXX Heisenberg model}
\label{subsec:XXX_solutions}

First, I state the result for isotropic XXX Heisenberg spin chain model with Hamiltonian
\begin{equation}
H^{XXX} = \frac{1}{2}\sum_{j=1}^L \left[ \mathbf{X}_j \mathbf{X}_{j+1} + \mathbf{Y}_j \mathbf{Y}_{j+1} + \mathbf{Z}_j \mathbf{Z}_{j+1} \right].
\end{equation}

\subsubsection{Integrals of motion}

There is a parametrized solution to Eq.~\eqref{eq:main_equation} which depends on three (possibly complex) variables $w,p,r$. The solution reads:
\begin{equation}
\label{eq:XXX_solution}
  \mathcal{A} =
  \begin{bmatrix}
    w^2+p^2+r^2 & p w\,\mathbf{X} & p w\,\mathbf{Y} & p w\,\mathbf{Z} \\
    p w\,\mathbf{X} & w^2 & -r w\,\mathbf{Z} & r w\,\mathbf{Y} \\
    p w\,\mathbf{Y} & r w\,\mathbf{Z} & w^2 & -r w\,\mathbf{X} \\
    p w\,\mathbf{Z} & -r w\,\mathbf{Y} & r w\,\mathbf{X} & w^2
  \end{bmatrix}
\end{equation}
with corresponding error term (defined up to additions $\mathcal{E}\mapsto\mathcal{E}+\gamma\mathcal{A}$)
\begin{equation}
\label{eq:XXX_error_term}
  \mathcal{E} =
  -
  \begin{bmatrix}
    2 r w & 0 & 0 & 0 \\
    0 & 2 r w & -(p^2+r^2)\,\mathbf{Z} & (p^2+r^2)\,\mathbf{Y} \\
    0 & (p^2+r^2)\,\mathbf{Z} & 2 r w & -(p^2+r^2)\,\mathbf{X} \\
    0 & -(p^2+r^2)\,\mathbf{Y} & (p^2+r^2)\,\mathbf{X} & 2 r w
  \end{bmatrix}
  .
\end{equation}
This solution generalizes one-parameter family of MPO charges found in \cite{Katsura_2015}, which is obtained by setting $r=0$. Kohei Fukai informed me that he also found these integrals of motion independently \cite{Private}.

\subsubsection{Spherical parametrization}

After examining this solution, it is natural to introduce spherical coordinates
\begin{equation}
\label{eq:spherical_coordinates}
  w = \rho \sin\theta, \quad p = \rho \cos\theta \cos\phi, \quad r = \rho \cos\theta \sin\phi.
\end{equation}
In terms of these coordinates, the solution reads
\begin{equation}
  \mathcal{A} =
  \rho^2
  \adjustbox{scale=0.9}{$
  \begin{bmatrix}
    1 & \sin\theta\cos\theta\cos\phi\,\mathbf{X} & \sin\theta\cos\theta\cos\phi\,\mathbf{Y} & \sin\theta\cos\theta\cos\phi\,\mathbf{Z} \\
    \sin\theta\cos\theta\cos\phi\,\mathbf{X} & \sin^2\theta & -\sin\theta\cos\theta\sin\phi\,\mathbf{Z} & \sin\theta\cos\theta\sin\phi\,\mathbf{Y} \\
    \sin\theta\cos\theta\cos\phi\,\mathbf{Y} & \sin\theta\cos\theta\sin\phi\,\mathbf{Z} & \sin^2\theta & -\sin\theta\cos\theta\sin\phi\,\mathbf{X} \\
    \sin\theta\cos\theta\cos\phi\,\mathbf{Z} & -\sin\theta\cos\theta\sin\phi\,\mathbf{Y} & \sin\theta\cos\theta\sin\phi\,\mathbf{X} & \sin^2\theta
  \end{bmatrix}
  $}
  .
\end{equation}
Here $\rho$ becomes a scaling parameter, which is insignificant. So, the solution is dependent on two-parameter family $\mathcal{A} = \mathcal{A}(\theta,\phi)$ where $(\theta,\phi)$ encode points of a sphere.

Multiplying $L$ local tensors $\mathcal{A}_j$ following Eq.~\eqref{eq:MPO_operator}, one constructs a two-parameter family of integrals of motion $\mathcal{M} = \mathcal{M}(\theta,\phi)$. At point $\theta=0$ and $\phi=0$ the charge is a unit operator $\mathcal{M}(0,0)=1$. Expanding $\mathcal{M}$ near the zero, the terms of expansion are local (that is, sums of close-range interactions) conserved quantities of the model:
\begin{equation}
  \mathcal{M}(\theta,\phi) = 1 + \theta^2\, H + \theta^3\phi\, H^{(3)} + \dots,
\end{equation}
where $H = H^{XXX}$ is a Hamiltonian and $H^{(3)}$ is a weight-$3$ local integral of motion \cite{Grabowski_1994}; higher terms in the series are also local charges by construction. I conjecture that $\mathcal{M}(\theta,\phi)$ is in fact complete for the class of local charges, meaning that it contains information about any local integral of motion of the XXX model, and any local charge can be obtained as a linear combination of coefficients of this series expansion. In comparison, the solution found in \cite{Katsura_2015,Fendley_2025} only contains information about even charges and corresponds to setting $\phi=0$.

Note that the spherical parametrization Eq.~\eqref{eq:spherical_coordinates} is not necessary the ``prettiest'', for example one might want to choose parameters $(\eta,\phi)$ where $e^\eta=\tan\theta$, so that:
\begin{equation}
  \mathcal{A} \propto
  \begin{bmatrix}
    e^\eta+e^{-\eta} & \cos\phi\,\mathbf{X} & \cos\phi\,\mathbf{Y} & \cos\phi\,\mathbf{Z} \\
    \cos\phi\,\mathbf{X} & e^\eta & -\sin\phi\,\mathbf{Z} & \sin\phi\,\mathbf{Y} \\
    \cos\phi\,\mathbf{Y} & \sin\phi\,\mathbf{Z} & e^\eta & -\sin\phi\,\mathbf{X} \\
    \cos\phi\,\mathbf{Z} & -\sin\phi\,\mathbf{Y} & \sin\phi\,\mathbf{X} & e^\eta
  \end{bmatrix}
  .
\end{equation}
Generally, I will consider different parametrizations of solutions later in the text.

\subsubsection{Trotterized case}

In case the dynamics of XXX spin chain is Trotterized with two alternating Hamiltonians acting during time $\tau$ each, a pair of operator matrices $\mathcal{A}_o$ and $\mathcal{A}_e$ that satisfy Eq.~\eqref{eq:trotterized_main_equation} is:
\begin{equation}
  \mathcal{A}_o = \mathcal{A}^{(0)}, \qquad \mathcal{A}_e = \mathcal{A}^{(0)} + \tan\tau\,\mathcal{A}^{(1)}+(\tan\tau)^2\,\mathcal{A}^{(2)},
\end{equation}
where I denote
\begin{equation}
\label{eq:XXX_floquet_solution}
\begin{aligned}
  \mathcal{A}^{(0)}
  &=
  \begin{bmatrix}
    w^2+p^2+r^2 & p w\,\mathbf{X} & p w\,\mathbf{Y} & p w\,\mathbf{Z} \\
    p w\,\mathbf{X} & w^2 & -r w\,\mathbf{Z} & r w\,\mathbf{Y} \\
    p w\,\mathbf{Y} & r w\,\mathbf{Z} & w^2 & -r w\,\mathbf{X} \\
    p w\,\mathbf{Z} & -r w\,\mathbf{Y} & r w\,\mathbf{X} & w^2
  \end{bmatrix}
  ,
  \\
  \mathcal{A}^{(1)}
  &=
  -
  \begin{bmatrix}
    2 r w & 0 & 0 & 0 \\
    0 & 2 r w & -(p^2+r^2)\,\mathbf{Z} & (p^2+r^2)\,\mathbf{Y} \\
    0 & (p^2+r^2)\,\mathbf{Z} & 2 r w & -(p^2+r^2)\,\mathbf{X} \\
    0 & -(p^2+r^2)\,\mathbf{Y} & (p^2+r^2)\,\mathbf{X} & 2 r w
  \end{bmatrix}
  ,
  \\
  \mathcal{A}^{(2)}
  &=
  \begin{bmatrix}
    p^2+r^2 & 0 & 0 & 0 \\
    0 & p^2+r^2 & 0 & 0 \\
    0 & 0 & p^2+r^2 & 0 \\
    0 & 0 & 0 & p^2+r^2
  \end{bmatrix}
  .
\end{aligned}
\end{equation}
As expected, in the limit $\tau\to 0$ both local tensors lead to solution Eq.~\eqref{eq:XXX_solution}, and the first derivative in $\tau$ equals to the error term Eq.~\eqref{eq:XXX_error_term}. That means, the solution Eq.~\eqref{eq:XXX_solution} is stable under Trotterizations.

This solution leads to a family of Floquet integrals of motion $\mathcal{M}_\tau$ in MPO form, it is dependent on two parameters $\mathcal{M}_\tau=\mathcal{M}_\tau(\theta,\phi)$, expanding $\mathcal{M}_\tau$ in the parameters near zero generates local Floquet charges of the model.

\subsection{XXZ Heisenberg model}
\label{subsec:XXZ_solutions}

Here I state the result for anisotropic XXZ Heisenberg spin chain model with Hamiltonian
\begin{equation}
H^{XXZ} = \frac{1}{2}\sum_{j=1}^L \left[\mathbf{X}_j\mathbf{X}_{j+1} + \mathbf{Y}_j \mathbf{Y}_{j+1} + \Delta\,\mathbf{Z}_j \mathbf{Z}_{j+1} \right].
\end{equation}
The solutions found for this model are generalizating XXX case solutions, but require more effort to find and to write down.

\subsubsection{Integrals of motion}

There is parametrized solution to Eq.~\eqref{eq:main_equation} which depends on three (possibly complex) variables $w,p,r$:
\begin{equation}
\label{eq:XXZ_solution}
  \mathcal{A} =
  \begin{bmatrix}
    \Delta(w^2+p^2+r^2) & \sqrt{\Delta}\, p w\,\mathbf{X} & \sqrt{\Delta}\, p w\,\mathbf{Y} & \sqrt{\Delta}\, p w\,\pi\,\mathbf{Z} \\
    \sqrt{\Delta}\, p w\,\mathbf{X} & w^2 & -r w\,\mathbf{Z} & r w \,\pi\,\mathbf{Y} \\
    \sqrt{\Delta}\, p w\,\mathbf{Y} & r w\,\mathbf{Z} & w^2 & -r w \,\pi\,\mathbf{X} \\
    \sqrt{\Delta}\, p w \,\pi\,\mathbf{Z} & -r w \,\pi\,\mathbf{Y} & r w \,\pi\,\mathbf{X} & w^2\,\omega
  \end{bmatrix}
  ,
\end{equation}
where I denote
\begin{equation}
  \pi = \sqrt{\Delta + \eta}, \qquad
  \eta= \left(\Delta-\Delta^{-1}\right) \frac{w^2}{p^2+r^2}, \qquad
  \omega = \frac{\Delta\, p^2+\Delta^{-1}r^2}{p^2+r^2}.
\end{equation}
The corresponding error term reads
\begin{equation}
\label{eq:XXZ_error_term}
  \mathcal{E} =
  \begin{bmatrix}
    -2 r w & \sqrt{\Delta}\, p r \,\eta\,\mathbf{X} & \sqrt{\Delta}\, p r \,\eta\,\mathbf{Y} & 0 \\
    \sqrt{\Delta}\, p r \,\eta\,\mathbf{X} & -2 r w & -[r^2\eta-(p^2+r^2)]\,\mathbf{Z} & -\Delta (p^2+r^2)\,\pi\,\mathbf{Y} \\
    \sqrt{\Delta}\, p r \,\eta\,\mathbf{Y} & [r^2\eta-(p^2+r^2)]\,\mathbf{Z} & -2 r w & \Delta (p^2+r^2)\,\pi\,\mathbf{X} \\
    0 & \Delta (p^2+r^2)\,\pi\,\mathbf{Y} & -\Delta (p^2+r^2)\,\pi\,\mathbf{X} & -2 r w
  \end{bmatrix}
  ,
\end{equation}
When $\Delta\to 1$, this solution replicates XXX case Eq.~\eqref{eq:XXX_solution}.

\subsubsection{Spherical parametrization}

Introducing spherical coordinates Eq.~\eqref{eq:spherical_coordinates}, the solution becomes
\begin{equation}
  \mathcal{A}
  = \rho^2
  \adjustbox{scale=0.8}{$
  \begin{bmatrix}
    \Delta & \sqrt{\Delta}\sin\theta\cos\theta\cos\phi\,\mathbf{X} & \sqrt{\Delta}\sin\theta\cos\theta\cos\phi\,\mathbf{Y} & \sqrt{\Delta}\sin\theta\cos\phi\,\pi\,\mathbf{Z} \\
    \sqrt{\Delta}\sin\theta\cos\theta\cos\phi\,\mathbf{X} & \sin^2\theta & -\sin\theta\cos\theta\sin\phi\,\mathbf{Z} & \sin\theta\sin\phi\,\pi\,\mathbf{Y} \\
    \sqrt{\Delta}\sin\theta\cos\theta\cos\phi\,\mathbf{Y} & \sin\theta\cos\theta\sin\phi\,\mathbf{Z} & \sin^2\theta & -\sin\theta\sin\phi\,\pi\,\mathbf{X} \\
    \sqrt{\Delta}\sin\theta\cos\phi\,\pi\,\mathbf{Z} & -\sin\theta\sin\phi\,\pi\,\mathbf{Y} & \sin\theta\sin\phi\,\pi\,\mathbf{X} & \sin^2\theta\,\omega
  \end{bmatrix}
  $},
\end{equation}
where
\begin{equation}
  \pi = \sqrt{\Delta-\Delta^{-1}\sin^2\theta}, \qquad
  \omega = \Delta\cos^2\phi+\Delta^{-1}\sin^2\phi.
\end{equation}
This solution leads to MPO integrals of motion $\mathcal{M}=\mathcal{M}(\theta,\phi)$ which depends on points of a sphere $(\theta,\phi)$. Local integrals of motion can be found by series expansion near $\theta=0$ and $\phi=0$. Analogously to the XXX case, I conjecture that such MPO family contains information about all local charges of the XXZ model.

Note that the presentation $\mathcal{M}$ as a sum of Pauli strings will not contain square roots such as $\sqrt{\Delta}$, because all non-diagonal terms appear twice in each term. That is why square roots are regularly used in this work.

In Section~\ref{subsec:XYZ_solutions}, I will mention another solution which lacks the inherent symmetry of the XXZ model.

\subsubsection{Diagonal elements parametrization}

Let me write down another parametrization of Eq.~\eqref{eq:XXZ_solution} which might appear less visual but turns out to be helpful. This form of solution depends on two parameters $\omega_x$ and $\omega_z$ sitting on the diagonal of the matrix and it reads:
\begin{equation}
\label{eq:XXZ_solution_diagonal}
  \mathcal{A} =
  \begin{bmatrix}
    1 & \scalebox{0.75}{$\sqrt{\frac{(\Delta\omega_x-1)(\omega_x-\Delta\omega_z)}{\Delta^2-1}}$}\,\mathbf{X} & \scalebox{0.75}{$\sqrt{\frac{(\Delta\omega_x-1)(\omega_x-\Delta\omega_z)}{\Delta^2-1}}$}\,\mathbf{Y} & \scalebox{0.75}{$\sqrt{\frac{(\omega_x-\Delta)(\omega_x-\Delta\omega_z)}{\Delta^2-1}}$}\,\mathbf{Z} \\
    \scalebox{0.75}{$\sqrt{\frac{(\Delta\omega_x-1)(\omega_x-\Delta\omega_z)}{\Delta^2-1}}$}\,\mathbf{X} & \omega_x & \scalebox{0.75}{$-\sqrt{\frac{(\Delta\omega_x-1)(\omega_z-\Delta\omega_x)}{\Delta^2-1}}$}\,\mathbf{Z} & \scalebox{0.75}{$\sqrt{\frac{(\omega_x-\Delta)(\omega_z-\Delta\omega_x)}{\Delta^2-1}}$}\,\mathbf{Y} \\
    \scalebox{0.75}{$\sqrt{\frac{(\Delta\omega_x-1)(\omega_x-\Delta\omega_z)}{\Delta^2-1}}$}\,\mathbf{Y} & \scalebox{0.75}{$\sqrt{\frac{(\Delta\omega_x-1)(\omega_z-\Delta\omega_x)}{\Delta^2-1}}$}\,\mathbf{Z} & \omega_x & \scalebox{0.75}{$-\sqrt{\frac{(\omega_x-\Delta)(\omega_z-\Delta\omega_x)}{\Delta^2-1}}$}\,\mathbf{X} \\
    \scalebox{0.75}{$\sqrt{\frac{(\omega_x-\Delta)(\omega_x-\Delta\omega_z)}{\Delta^2-1}}$}\,\mathbf{Z} & \scalebox{0.75}{$-\sqrt{\frac{(\omega_x-\Delta)(\omega_z-\Delta\omega_x)}{\Delta^2-1}}$}\,\mathbf{Y} & \scalebox{0.75}{$\sqrt{\frac{(\omega_x-\Delta)(\omega_z-\Delta\omega_x)}{\Delta^2-1}}$}\,\mathbf{X} & \omega_z
  \end{bmatrix}
  ,
\end{equation}
the error term is
\begin{equation}
\label{eq:XXZ_error_term_2}
  \mathcal{E} =
  \scalebox{0.85}{$
  -
  \begin{bmatrix}
    \scalebox{0.8}{$2\sqrt{\frac{(\Delta\omega_x-1)(\omega_z-\Delta\omega_x)}{\Delta^2-1}}$} & \scalebox{0.7}{$\sqrt{(\omega_z-\Delta\omega_x)(\omega_x-\Delta\omega_z)}$}\,\mathbf{X} & \scalebox{0.7}{$\sqrt{(\omega_z-\Delta\omega_x)(\omega_x-\Delta\omega_z)}$}\,\mathbf{Y} & 0 \\
    \scalebox{0.7}{$\sqrt{(\omega_z-\Delta\omega_x)(\omega_x-\Delta\omega_z)}$}\,\mathbf{X} & \scalebox{0.8}{$2 \Delta\sqrt{\frac{(\Delta\omega_x-1)(\omega_z-\Delta\omega_x)}{\Delta^2-1}}$} & \scalebox{0.8}{$-(1+\omega_z-2\Delta\omega_x)$}\,\mathbf{Z} & \scalebox{0.7}{$\sqrt{(\omega_x-\Delta)(\Delta\omega_x-1)}$}\,\mathbf{Y} \\
    \scalebox{0.7}{$\sqrt{(\omega_z-\Delta\omega_x)(\omega_x-\Delta\omega_z)}$}\,\mathbf{Y} & \scalebox{0.8}{$(1+\omega_z-2\Delta\omega_x)$}\,\mathbf{Z} & \scalebox{0.5}{$2 \Delta\sqrt{\frac{(\Delta\omega_x-1)(\omega_z-\Delta\omega_x)}{\Delta^2-1}}$} & \scalebox{0.7}{$-\sqrt{(\omega_x-\Delta)(\Delta\omega_x-1)}$}\,\mathbf{X} \\
    0 & \scalebox{0.7}{$-\sqrt{(\omega_x-\Delta)(\Delta\omega_x-1)}$}\,\mathbf{Y} & \scalebox{0.7}{$\sqrt{(\omega_x-\Delta)(\Delta\omega_x-1)}$}\,\mathbf{X} & \scalebox{0.8}{$2\sqrt{\frac{(\Delta\omega_x-1)(\omega_z-\Delta\omega_x)}{\Delta^2-1}}$}
  \end{bmatrix}
  $}
  .
\end{equation}
This form of solution will be used for reference later in the text.

\subsubsection{Trotterized case}

In case the dynamics of XXZ spin chain is Trotterized with two alternating Hamiltonians acting for time period $\tau$, a pair of operator matrices $\mathcal{A}_o$ and $\mathcal{A}_e$ that satisfy Eq.~\eqref{eq:trotterized_main_equation} is:
\begin{equation}
\label{eq:XXZ_floquet_solution}
\begin{aligned}
  \mathcal{A}_o &=
  \begin{bmatrix}
    \Delta_\tau(w^2+p^2+r^2) & \sqrt{\Delta_\tau}\, p w\,\mathbf{X} & \sqrt{\Delta_\tau}\, p w\,\mathbf{Y} & \sqrt{\Delta_\tau}\, p w\,\pi_\tau\,\mathbf{Z} \\
    \sqrt{\Delta_\tau}\, p w\,\mathbf{X} & w^2 & -r w\,\mathbf{Z} & r w \,\pi_\tau\,\mathbf{Y} \\
    \sqrt{\Delta_\tau}\, p w\,\mathbf{Y} & r w\,\mathbf{Z} & w^2 & -r w \,\pi_\tau\,\mathbf{X} \\
    \sqrt{\Delta_\tau}\, p w \,\pi_\tau\,\mathbf{Z} & -r w \,\pi_\tau\,\mathbf{Y} & r w \,\pi_\tau\,\mathbf{X} & w^2\,\frac{\Delta_\tau p^2+\Delta_\tau^{-1}r^2}{p^2+r^2}
  \end{bmatrix}
  ,
  \\
  \mathcal{A}_e &=
  \begin{bmatrix}
    \nu & \sqrt{\Delta_\tau}\, p w\,\psi\,\mathbf{X} & \sqrt{\Delta_\tau}\, p w\,\psi\,\mathbf{Y} & \sqrt{\Delta_\tau}\, p w\,\pi_\tau\,\mathbf{Z} \\
    \sqrt{\Delta_\tau}\, p w\,\psi\,\mathbf{X} & w^2\,\frac{p^2+r_\tau^2}{p^2+r^2} & -r_\tau w\,\psi\,\mathbf{Z} & r_\tau w \,\pi_\tau\,\mathbf{Y} \\
    \sqrt{\Delta_\tau}\, p w\,\psi\,\mathbf{Y} & r_\tau w\,\psi\,\mathbf{Z} & w^2\,\frac{p^2+r_\tau^2}{p^2+r^2} & -r_\tau w \,\pi_\tau\,\mathbf{X} \\
    \sqrt{\Delta_\tau}\, p w\,\pi_\tau\,\mathbf{Z} & -r_\tau w \,\pi_\tau\,\mathbf{Y} & r_\tau w \,\pi_\tau\,\mathbf{X} & w^2\,\frac{\Delta_\tau p^2+\Delta_\tau^{-1}r_\tau^2}{p^2+r^2}
  \end{bmatrix}
  ,
\end{aligned}
\end{equation}
where I denote
\begin{equation}
\begin{aligned}
  \nu &= \Delta_\tau(w^2+p^2+r^2)\,\psi - 2 r w\,\psi\,\tan\tau + \left[\Delta_\tau(p^2+r^2)+r^2\eta_\tau\right]\,(\tan\tau)^2, \\
  \psi &= \frac{\cos(\Delta\tau)}{\cos\tau}+\frac{r}{w}\,\eta_\tau\tan\tau, \\
  r_\tau &= r\,\frac{\cos(\Delta\tau)}{\cos\tau} - \Delta_\tau\,\frac{p^2+r^2}{w}\,\tan\tau,\\
  \pi_\tau &= \sqrt{\Delta_\tau + \eta_\tau}, \\
  \eta_\tau &= \left(\Delta_\tau-\Delta_\tau^{-1}\right) \frac{w^2}{p^2+r^2}, \\
  \Delta_\tau &= \frac{\sin(\Delta\tau)}{\sin\tau}.
\end{aligned}
\end{equation}
One can show that in the limit $\tau\to 0$ both local tensors lead to solution Eq.~\eqref{eq:XXX_solution}, and the first derivative in $\tau$ equals to the error term Eq.~\eqref{eq:XXZ_error_term}; in the limit $\Delta\to 1$ the solution reproduces XXX Floquet charges Eq.~\eqref{eq:XXX_floquet_solution}.

This solution leads to a family of Floquet integrals of motion $\mathcal{M}$ in MPO form dependent on two parameters, expanding $\mathcal{M}$ in the parameters near zero generates local Floquet charges of the Trotterized XXZ Heisenberg model.

\subsection{XX Heisenberg model}

Here I write down the solution to XX Heisenberg model with Hamiltonian
\begin{equation}
  H^{XX} = \frac{1}{2}\sum_{j=1}^L \left[ \mathbf{X}_j\mathbf{X}_{j+1} + \mathbf{Y}_j\mathbf{Y}_{j+1} \right].
\end{equation}
This model is free-fermionic and can be solved using more elegant methods, its local integrals of motion are easy to describe for example in terms of Onsager strings \cite{Perk_2017,Lychkovskiy_2021}. Still, the XX charges (and XY charges from the next subsection) comprise a good consistency check for the method.

\subsubsection{Integrals of motion}

Taking a limit $\Delta\to 0$ in the XXZ solution Eq.~\eqref{eq:XXZ_solution_diagonal}, one obtains XX conserved quantities as
\begin{equation}
\label{eq:XX_solution}
  \mathcal{A} =
  \begin{bmatrix}
    1 & \sqrt{\omega_x}\,\mathbf{X} & \sqrt{\omega_x}\,\mathbf{Y} & i\,\omega_x\,\mathbf{Z} \\
    \sqrt{\omega_x}\,\mathbf{X} & \omega_x & -\sqrt{\omega_z}\,\mathbf{Z} & i\,\sqrt{\omega_x\omega_z}\,\mathbf{Y} \\
    \sqrt{\omega_x}\,\mathbf{Y} & \sqrt{\omega_z}\,\mathbf{Z} & \omega_x & -i\,\sqrt{\omega_x\omega_z}\,\mathbf{X} \\
    i\,\omega_x\,\mathbf{Z} & -i\,\sqrt{\omega_x\omega_z}\,\mathbf{Y} & i\,\sqrt{\omega_x\omega_z}\,\mathbf{X} & \omega_z
  \end{bmatrix}
  ,
\end{equation}
where $\omega_x,\omega_z$ are two parameters and $i$ is the imaginary unit. The corresponding error term is
\begin{equation}
  \mathcal{E} =
  -
  \begin{bmatrix}
    2\sqrt{\omega_z} & \sqrt{\omega_x\omega_z}\,\mathbf{X} & \sqrt{\omega_x\omega_z}\,\mathbf{Y} & 0 \\
    \sqrt{\omega_x\omega_z}\,\mathbf{X} & 0 & -(1+\omega_z)\,\mathbf{Z} & i\sqrt{\omega_x}\,\mathbf{Y} \\
    \sqrt{\omega_x\omega_z}\,\mathbf{Y} & (1+\omega_z)\,\mathbf{Z} & 0 & -i\sqrt{\omega_x}\,\mathbf{X} \\
    0 & -i\sqrt{\omega_x}\,\mathbf{Y} & i\sqrt{\omega_x}\,\mathbf{X} & 2\sqrt{\omega_z}
  \end{bmatrix}
  .
\end{equation}
One can check that when expanding the resulting MPO into series near zero $\omega_x=0$ and $\omega_z=0$, the series coefficients are products of Onsager strings and strings made of $Z$ operators with even weight, which supports the conjecture of completeness of the MPO charges.

\subsubsection{Trotterized case}

For Trotterized XX model, the operator matrices $\mathcal{A}_o$ and $\mathcal{A}_e$ satisfying Eq.~\eqref{eq:trotterized_main_equation} are
\begin{equation}
\label{eq:XX_floquet_solution}
\begin{aligned}
  \mathcal{A}_o &=
  \begin{bmatrix}
    1 & \sqrt{\omega_x}\,\mathbf{X} & \sqrt{\omega_x}\,\mathbf{Y} & i\,\omega_x\,\mathbf{Z} \\
    \sqrt{\omega_x}\,\mathbf{X} & \omega_x & -\sqrt{\omega_z}\,\mathbf{Z} & i\,\sqrt{\omega_x\omega_z}\,\mathbf{Y} \\
    \sqrt{\omega_x}\,\mathbf{Y} & \sqrt{\omega_z}\,\mathbf{Z} & \omega_x & -i\,\sqrt{\omega_x\omega_z}\,\mathbf{X} \\
    i\,\omega_x\,\mathbf{Z} & -i\,\sqrt{\omega_x\omega_z}\,\mathbf{Y} & i\,\sqrt{\omega_x\omega_z}\,\mathbf{X} & \omega_z
  \end{bmatrix}
  , \\
  \mathcal{A}_e &=
  \begin{bmatrix}
    1 & \sqrt{\omega^\tau_x}\,\mathbf{X} & \sqrt{\omega^\tau_x}\,\mathbf{Y} & i\,\omega^\tau_x\,\mathbf{Z} \\
    \sqrt{\omega^\tau_x}\,\mathbf{X} & \omega^\tau_x & -\sqrt{\omega^\tau_z}\,\mathbf{Z} & i\,\sqrt{\omega^\tau_x\omega^\tau_z}\,\mathbf{Y} \\
    \sqrt{\omega^\tau_x}\,\mathbf{Y} & \sqrt{\omega^\tau_z}\,\mathbf{Z} & \omega^\tau_x & -i\,\sqrt{\omega^\tau_x\omega^\tau_z}\,\mathbf{X} \\
    i\,\omega^\tau_x\,\mathbf{Z} & -i\,\sqrt{\omega^\tau_x\omega^\tau_z}\,\mathbf{Y} & i\,\sqrt{\omega^\tau_x\omega^\tau_z}\,\mathbf{X} & \omega^\tau_z
  \end{bmatrix}
  ,
\end{aligned}
\end{equation}
where
\begin{equation}
  \sqrt{\omega_x^\tau} = \frac{\cos\tau\sqrt{\omega_x}}{1-\sin\tau\,\sqrt{\omega_z}}, \qquad
  \sqrt{\omega_z^\tau} = \frac{\sqrt{\omega_z}-\sin\tau}{1-\sin\tau\,\sqrt{\omega_z}}.
\end{equation}

\subsection{XY Heisenberg model}

Here I write down the solution to XY Heisenberg model with Hamiltonian
\begin{equation}
  H^{XY} = \frac{1}{2}\sum_{j=1}^L \left[ J_x\,\mathbf{X}_j\mathbf{X}_{j+1} + J_y\,\mathbf{Y}_j\mathbf{Y}_{j+1} \right].
\end{equation}

\subsubsection{Integrals of motion}

Generalizing Eq.~\eqref{eq:XX_solution}, one finds a solution to the XY Heisenberg model:
\begin{equation}
\label{eq:XY_solution}
  \mathcal{A} =
  \begin{bmatrix}
    1 & \sqrt{\omega_x}\,\mathbf{X} & \sqrt{\omega_y}\,\mathbf{Y} & i \sqrt{\omega_x\omega_y}\,\mathbf{Z} \\
    \sqrt{\omega_x}\,\mathbf{X} & \omega_x & -\sqrt{\omega_z}\,\mathbf{Z} & i\sqrt{\omega_x\omega_z}\,\mathbf{Y} \\
    \sqrt{\omega_y}\,\mathbf{Y} & \sqrt{\omega_z}\,\mathbf{Z} & \omega_y & -i\sqrt{\omega_y\omega_z}\,\mathbf{X} \\
    i \sqrt{\omega_x\omega_y}\,\mathbf{Z} & -i\sqrt{\omega_x\omega_z}\,\mathbf{Y} & i\sqrt{\omega_y\omega_z}\,\mathbf{X} & \omega_z
  \end{bmatrix}
  ,
\end{equation}
where now $\omega_x,\omega_y$ are two parameters and $\omega_z$ is expressed in terms of them as
\begin{equation}
  \omega_z = \frac{J_y\omega_x-J_x\omega_y}{J_x\omega_x-J_y\omega_y}.
\end{equation}
The corresponding error term is
\begin{equation}
  \mathcal{E} =
  -\scalebox{0.9}{$\sqrt{\frac{\omega_z}{\omega_x\omega_y}}$}
  \begin{bmatrix}
    \scalebox{0.9}{$J_x\omega_x+J_y\omega_y$} & J_x\sqrt{\omega_x}^3\,\mathbf{X} & J_y\sqrt{\omega_y}^3\,\mathbf{Y} & 0 \\
    J_x\sqrt{\omega_x}^3\,\mathbf{X} & \scalebox{0.9}{$(J_x\omega_x-J_y\omega_y)\,\omega_x$} & -\frac{J_xJ_y(\omega_x^2-\omega_y^2)}{\sqrt{\omega_z}(J_x\omega_x-J_y\omega_y)}\,\mathbf{Z} & -iJ_y\frac{\sqrt{\omega_x}^3}{\sqrt{\omega_z}}\,\mathbf{Y} \\
    J_y\sqrt{\omega_y}^3\,\mathbf{Y} & \frac{J_xJ_y(\omega_x^2-\omega_y^2)}{\sqrt{\omega_z}(J_x\omega_x-J_y\omega_y)}\,\mathbf{Z} & \scalebox{0.9}{$(J_y\omega_y-J_x\omega_x)\,\omega_y$} & iJ_x\frac{\sqrt{\omega_y}^3}{\sqrt{\omega_z}}\,\mathbf{X} \\
    0 & -iJ_y\frac{\sqrt{\omega_x}^3}{\sqrt{\omega_z}}\,\mathbf{Y} & iJ_x\frac{\sqrt{\omega_y}^3}{\sqrt{\omega_z}}\,\mathbf{X} & \scalebox{0.9}{$J_x\omega_x+J_y\omega_y$}
  \end{bmatrix}
  .
\end{equation}
In the XX limit $J_x\to 1$, $J_y\to 1$, this solution aligns with Eq.~\eqref{eq:XX_solution}.

\subsubsection{Trotterized case}

For Trotterized XY model, the operator matrices $\mathcal{A}_o$ and $\mathcal{A}_e$ satisfying Eq.~\eqref{eq:trotterized_main_equation} are
\begin{equation}
\label{eq:XY_floquet_solution}
\begin{aligned}
  \mathcal{A}_o =
  \begin{bmatrix}
    1 & \sqrt{\omega_x}\,\mathbf{X} & \sqrt{\omega_y}\,\mathbf{Y} & i \sqrt{\omega_x\omega_y}\,\mathbf{Z} \\
    \sqrt{\omega_x}\,\mathbf{X} & \omega_x & -\sqrt{\omega_z}\,\mathbf{Z} & i\sqrt{\omega_x\omega_z}\,\mathbf{Y} \\
    \sqrt{\omega_y}\,\mathbf{Y} & \sqrt{\omega_z}\,\mathbf{Z} & \omega_y & -i\sqrt{\omega_y\omega_z}\,\mathbf{X} \\
    i \sqrt{\omega_x\omega_y}\,\mathbf{Z} & -i\sqrt{\omega_x\omega_z}\,\mathbf{Y} & i\sqrt{\omega_y\omega_z}\,\mathbf{X} & \omega_z
  \end{bmatrix}
  ,\\
  \mathcal{A}_e =
  \begin{bmatrix}
    1 & \sqrt{\omega^\tau_x}\,\mathbf{X} & \sqrt{\omega^\tau_y}\,\mathbf{Y} & i \sqrt{\omega^\tau_x\omega^\tau_y}\,\mathbf{Z} \\
    \sqrt{\omega^\tau_x}\,\mathbf{X} & \omega^\tau_x & -\sqrt{\omega^\tau_z}\,\mathbf{Z} & i\sqrt{\omega^\tau_x\omega^\tau_z}\,\mathbf{Y} \\
    \sqrt{\omega^\tau_y}\,\mathbf{Y} & \sqrt{\omega^\tau_z}\,\mathbf{Z} & \omega^\tau_y & -i\sqrt{\omega^\tau_y\omega^\tau_z}\,\mathbf{X} \\
    i \sqrt{\omega^\tau_x\omega^\tau_y}\,\mathbf{Z} & -i\sqrt{\omega^\tau_x\omega^\tau_z}\,\mathbf{Y} & i\sqrt{\omega^\tau_y\omega^\tau_z}\,\mathbf{X} & \omega^\tau_z
  \end{bmatrix}
  ,
\end{aligned}
\end{equation}
where I denote
\begin{equation}
\begin{aligned}
  \omega_z &= \frac{\sin(J_y\tau)\,\omega_x-\sin(J_x\tau)\,\omega_y}{\sin(J_x\tau)\,\omega_x-\sin(J_y\tau)\,\omega_y},\\
  \sqrt{\omega_x^\tau} &= \delta \left[\cos(J_y\tau)\sqrt{\omega_x}+\cos(J_x\tau)\sin(J_y\tau)\sqrt{\omega_y\omega_z}\right], \\
  \sqrt{\omega_y^\tau} &= \delta \left[\cos(J_x\tau)\sqrt{\omega_y}+\cos(J_y\tau)\sin(J_x\tau)\sqrt{\omega_x\omega_z}\right], \\
  \sqrt{\omega_z^\tau} &= \delta \left[\cos(J_x\tau)\cos(J_y\tau)\sqrt{\omega_z}+\frac{\cos(J_x\tau)^2-\cos(J_y\tau)^2}{\sin(J_x\tau)\omega_x-\sin(J_y\tau)\omega_y}\sqrt{\omega_x\omega_y}\right], \\
  \delta &= \frac{\sin(J_x\tau)\,\omega_x-\sin(J_y\tau)\,\omega_y}{[\cos(J_y\tau)]^2\sin(J_x\tau)\,\omega_x-[\cos(J_x\tau)]^2\sin(J_y\tau)\,\omega_y}.
\end{aligned}
\end{equation}

\subsection{XYZ Heisenberg model}
\label{subsec:XYZ_solutions}

Now, consider the general anisotropic Heisenberg spin chain model with Hamiltonian
\begin{equation}
  H^{XYZ} = \frac{1}{2}\sum_{j=1}^L \left[ J_x\,\mathbf{X}_j\mathbf{X}_{j+1} + J_y\,\mathbf{Y}_j\mathbf{Y}_{j+1} + J_z\,\mathbf{Z}_j\mathbf{Z}_{j+1} \right].
\end{equation}
Here I describe a two-parameter family of MPO integrals of motion with bond dimension $4$, which generalizes all the solutions listed above.

\subsubsection{Integrals of motion}

There is a solution to Eq.~\eqref{eq:main_equation} that depends on (generally complex) parameters $\pi_x,\pi_y,\pi_z$ and is given by
\begin{equation}
\label{eq:XYZ_solution}
  \mathcal{A} =
  \begin{bmatrix}
  1 & \sqrt{\pi_x\frac{\Pi_2}{\Pi_3}}\,\mathbf{X} & \sqrt{\pi_y\frac{\Pi_2}{\Pi_3}}\,\mathbf{Y} & \sqrt{\pi_z\frac{\Pi_2}{\Pi_3}}\,\mathbf{Z} \\
  \sqrt{\pi_x\frac{\Pi_2}{\Pi_3}}\,\mathbf{X} & \omega_x & -\sqrt{\pi_x\pi_y\frac{\Pi_1}{\Pi_3}}\,\mathbf{Z} & \sqrt{\pi_z\pi_x\frac{\Pi_1}{\Pi_3}}\,\mathbf{Y} \\
  \sqrt{\pi_y\frac{\Pi_2}{\Pi_3}}\,\mathbf{Y} & \sqrt{\pi_x\pi_y\frac{\Pi_1}{\Pi_3}}\,\mathbf{Z} & \omega_y & -\sqrt{\pi_y\pi_z\frac{\Pi_1}{\Pi_3}}\,\mathbf{X} \\
  \sqrt{\pi_z\frac{\Pi_2}{\Pi_3}}\,\mathbf{Z} & -\sqrt{\pi_z\pi_x\frac{\Pi_1}{\Pi_3}}\,\mathbf{Y} & \sqrt{\pi_y\pi_z\frac{\Pi_1}{\Pi_3}}\,\mathbf{X} & \omega_z
  \end{bmatrix}
  ,
\end{equation}
where abbreviations $\Pi_1$,$\Pi_2$,$\Pi_3$ descibe the determinants
\begin{equation}
\begin{gathered}
  \Pi_1 = \Det\begin{bmatrix}1&1&1\\ J_x^2&J_y^2&J_z^2\\ J_x\pi_x&J_y\pi_y&J_z\pi_z\end{bmatrix}, \qquad
  \Pi_2 = \Det\begin{bmatrix}1&1&1\\ J_x^2&J_y^2&J_z^2\\ J_x\pi_y\pi_z&J_y\pi_z\pi_x&J_z\pi_x\pi_y\end{bmatrix},
  \\
  \Pi_3 = \Det\begin{bmatrix}1&1&1\\ J_x\pi_x&J_y\pi_y&J_z\pi_z\\ J_x^2\pi_x^2&J_y^2\pi_y^2&J_z^2\pi_z^2\end{bmatrix}
  = (J_y\pi_y-J_z\pi_z)(J_z\pi_z-J_x\pi_x)(J_x\pi_x-J_y\pi_y).
\end{gathered}
\end{equation}
and the diagonal elements $\omega_x,\omega_y,\omega_z$ are expressed from $\pi_x,\pi_y,\pi_z$ as
\begin{equation}
  \omega_x = \frac{J_z\pi_y-J_y\pi_z}{J_y\pi_y-J_z\pi_z}, \qquad
  \omega_y = \frac{J_x\pi_z-J_z\pi_x}{J_z\pi_z-J_x\pi_x}, \qquad
  \omega_z = \frac{J_y\pi_x-J_x\pi_y}{J_x\pi_x-J_y\pi_y}.
\end{equation}

The error term (defined up to additions $\mathcal{E}\mapsto \mathcal{E}+\gamma\mathcal{A}$) corresponding to Eq~\eqref{eq:XYZ_solution} is
\begin{equation}
  \mathcal{E} =
  \adjustbox{scale=0.8}{$
  \begin{bmatrix}
  0 & \sqrt{\pi_x \Pi_1\Pi_2} \frac{J_y\pi_y+J_z\pi_z}{\Pi_3}\,\mathbf{X} & \sqrt{\pi_y \Pi_1\Pi_2} \frac{J_z\pi_z+J_x\pi_x}{\Pi_3}\,\mathbf{Y} & \sqrt{\pi_z \Pi_1\Pi_2} \frac{J_x\pi_x+J_y\pi_y}{\Pi_3}\,\mathbf{Z} \\
  \sqrt{\pi_x \Pi_1\Pi_2} \frac{J_y\pi_y+J_z\pi_z}{\Pi_3}\,\mathbf{X} & -2\sqrt{\frac{\Pi_1}{\Pi_3}}\frac{(J_y^2-J_z^2)\pi_y\pi_z}{J_y\pi_y-J_z\pi_z} & -\sqrt{\pi_x\pi_y}\frac{\widetilde{\Pi}_2-J_z\pi_z\Pi_1}{\Pi_3}\,\mathbf{Z} & \sqrt{\pi_z\pi_x}\frac{\widetilde{\Pi}_2-J_y\pi_y\Pi_1}{\Pi_3}\,\mathbf{Y} \\
  \sqrt{\pi_y \Pi_1\Pi_2} \frac{J_z\pi_z+J_x\pi_x}{\Pi_3}\,\mathbf{Y} & \sqrt{\pi_x\pi_y}\frac{\widetilde{\Pi}_2-J_z\pi_z\Pi_1}{\Pi_3}\,\mathbf{Z} & -2\sqrt{\frac{\Pi_1}{\Pi_3}}\frac{(J_z^2-J_x^2)\pi_z\pi_x}{J_z\pi_z-J_x\pi_x} & -\sqrt{\pi_y\pi_z}\frac{\widetilde{\Pi}_2-J_x\pi_x\Pi_1}{\Pi_3}\,\mathbf{X} \\
  \sqrt{\pi_z \Pi_1\Pi_2} \frac{J_x\pi_x+J_y\pi_y}{\Pi_3}\,\mathbf{Z} & -\sqrt{\pi_z\pi_x}\frac{\widetilde{\Pi}_2-J_y\pi_y\Pi_1}{\Pi_3}\,\mathbf{Y} & \sqrt{\pi_y\pi_z}\frac{\widetilde{\Pi}_2-J_x\pi_x\Pi_1}{\Pi_3}\,\mathbf{X} & -2\sqrt{\frac{\Pi_1}{\Pi_3}}\frac{(J_x^2-J_y^2)\pi_x\pi_y}{J_x\pi_x-J_y\pi_y}
  \end{bmatrix}
  $},
\end{equation}
where
\begin{equation}
  \widetilde{\Pi}_2 = \Det\begin{bmatrix}1&1&1\\ J_x\pi_x&J_y\pi_y&J_z\pi_z\\ J_x^3\pi_x&J_y^3\pi_y&J_z^3\pi_z\end{bmatrix}.
\end{equation}
This solution defines a set of MPO charges $\mathcal{M}=\mathcal{M}(\pi_x,\pi_y,\pi_z)$ which is invariant under parameter scaling $(\pi_x,\pi_y,\pi_z)\mapsto(\lambda \pi_x,\lambda\pi_y,\lambda\pi_z)$, it is natural to consider it as depending on homogeneous coordinates $[\pi_x\!:\!\pi_y\!:\!\pi_z]$ describing points of (complex) projective plane. One can try to generate local charges by choosing zero point $\omega_x=\omega_y=\omega_z=0$ (next subsection explains how) and doing series expansion. Once again, I conjecture that all local charges are generated in this way, but proving it for XYZ case should be difficult.

\subsubsection{Limits of the solution}

The solution Eq.~\eqref{eq:XYZ_solution} is depends on two-dimensional space of parameters. Depending on the context, one should consider different parametrizations, which may lead to different presentations of solution.

One important reparametrization is two choose two variables from $\omega_x,\omega_y,\omega_z$. The elements $\omega_x,\omega_y,\omega_z$ satisfy algebraic equation
\begin{equation}
  \Det\begin{bmatrix}1&1&1\\ J_x^2&J_y^2&J_z^2\\ J_x(\omega_x+\omega_y\omega_z)&J_y(\omega_y+\omega_z\omega_x)&J_z(\omega_z+\omega_x\omega_y)\end{bmatrix}
  = 0.
\end{equation}
When fixing $\omega_z$ and $\omega_x$, this equation is linear in parameter $\omega_y$, so $\omega_y$ can be expressed as a rational expression of $\omega_z,\omega_x$ (similarly for cyclic shifts of $x,y,z$):
\begin{equation}
  \omega_y = \frac{J_y(J_z^2-J_x^2)\,\omega_x\omega_z+J_x(J_y^2-J_z^2)\,\omega_x+J_z(J_x^2-J_y^2)\,\omega_z}{J_y(J_x^2-J_z^2)+J_z(J_y^2-J_x^2)\,\omega_x+J_x(J_z^2-J_y^2)\,\omega_z}.
\end{equation}
Also, the elements $\pi_x,\pi_z$ can be expressed from $\omega_x,\omega_z$ and $\pi_y$ as
\begin{equation}
  \pi_z = \pi_y \frac{J_z-J_y\omega_x}{J_y-J_z\omega_x}, \qquad
  \pi_x = \pi_y \frac{J_x-J_y\omega_z}{J_y-J_x\omega_z}.
\end{equation}
After such substitution the parameter $\pi_y$ is vanishes because of scaling invariance, so the solution Eq.~\eqref{eq:XYZ_solution} becomes an expression of $\omega_z, \omega_x$. Note that at zero $\omega_z=\omega_x=0$ the MPO is trivial $\mathcal{M}=1$, and one can try to expand $\mathcal{M}$ near this point. After choosing parameters $\omega_x$ and $\omega_z$, taking the XXZ limit $J_x\to 1$, $J_y\to 1$, $J_z\to \Delta$ leads directly to solution Eq.~\eqref{eq:XXZ_solution_diagonal}. Also, taking the XY limit $J_z\to 0$, one easily finds XY solution Eq.~\eqref{eq:XY_solution}.

On the other hand, choosing incorrect parametrization may be harmful. When taking the XXZ limit using parameters $\pi_x,\pi_y,\pi_z$ from Eq.~\eqref{eq:XYZ_solution}, it becomes
\begin{equation}
\label{eq:XXZ_plane_solution}
  \mathcal{A} =
  \begin{bmatrix}
  1 & \sqrt{\frac{\pi_x}{\Pi}}\mathbf{X} & \sqrt{\frac{\pi_y}{\Pi}}\,\mathbf{Y} & \sqrt{\frac{1}{\Pi}}\,\mathbf{Z} \\
  \sqrt{\frac{\pi_x}{\Pi}}\,\mathbf{X} & \frac{\Delta\pi_y-1}{\pi_y-\Delta} & -\sqrt{-\frac{\pi_x\pi_y}{\Pi}}\,\mathbf{Z} & \sqrt{-\frac{\pi_x}{\Pi}}\,\mathbf{Y} \\
  \sqrt{\frac{\pi_y}{\Pi}}\,\mathbf{Y} & \sqrt{-\frac{\pi_x\pi_y}{\Pi}}\,\mathbf{Z} & \frac{\Delta\pi_x-1}{\pi_x-\Delta} & -\sqrt{-\frac{\pi_y}{\Pi}}\,\mathbf{X} \\
  \sqrt{\frac{1}{\Pi}}\,\mathbf{Z}      & -\sqrt{-\frac{\pi_x}{\Pi}}\,\mathbf{Y} & \sqrt{-\frac{\pi_y}{\Pi}}\,\mathbf{X} & 1\\
  \end{bmatrix}
  ,
\end{equation}
where I set $\pi_z=1$ and denote
\begin{equation}
  \Pi=\frac{(\pi_x-\Delta)(\pi_y-\Delta)}{\Delta^2-1}.
\end{equation}
The drawback of this XXZ solution is that the tensor $\mathcal{A}$ is not symmetric under rotations over $\mathbf{Z}$. I will formulate what I mean by symmetry invariance in Section~\ref{subsec:symmetry}. In the XXX limit $\Delta\to 1$, this operator matrix $\mathcal{A}$ tends to identity matrix and thus becomes trivial; in the XX limit $\Delta\to 0$ it is consistent with Eq.~\eqref{eq:XX_solution}.

\subsubsection{Two one-parameter solutions}

There two known one-parameter families of conserved quantities of XYZ model. They turn out to be the limiting cases of Eq.~\eqref{eq:XYZ_solution}.

The first one-parameter family is the solution of \cite{Fendley_2025}, which is dependent on a parameter $\zeta$ and reads
\begin{equation}
  \mathcal{A} =
  \begin{bmatrix}
    1 & \scalebox{0.8}{$\sqrt{\zeta(J_x-J_yJ_z\zeta)}$}\,\mathbf{X} & \scalebox{0.8}{$\sqrt{\zeta(J_y-J_zJ_x\zeta)}$}\,\mathbf{Y} & \scalebox{0.8}{$\sqrt{\zeta(J_z-J_xJ_y\zeta)}$}\,\mathbf{Z} \\
    \scalebox{0.8}{$\sqrt{\zeta(J_x-J_yJ_z\zeta)}$}\,\mathbf{X} & J_x \zeta & 0 & 0 \\
    \scalebox{0.8}{$\sqrt{\zeta(J_y-J_zJ_x\zeta)}$}\,\mathbf{Y} & 0 & J_y \zeta & 0 \\
    \scalebox{0.8}{$\sqrt{\zeta(J_z-J_xJ_y\zeta)}$}\,\mathbf{Z} & 0 & 0 & J_z \zeta \\
  \end{bmatrix}
  .
\end{equation}
This solution can be obtained from Eq.~\eqref{eq:XYZ_solution} after taking the limit
\begin{equation}
  \omega_x \to J_x\zeta, \qquad
  \omega_y \to J_y\zeta, \qquad
  \omega_z \to J_z\zeta.
\end{equation}

The other one-parameter solution, which was found in \cite[Eq.~(45)]{Fukai_2026} in slightly different form, can be obtained in the limit
\begin{equation}
  \omega_x \to \frac{\zeta}{J_x}, \qquad
  \omega_y \to \frac{\zeta}{J_y}, \qquad
  \omega_z \to \frac{\zeta}{J_z},
\end{equation}
and it reads
\begin{equation}
  \mathcal{A} =
  \begin{bmatrix}
    1 & 0 & 0 & 0 \\
    0 & \frac{\zeta}{J_x} & -\frac{\sqrt{\zeta(\zeta-J_z^2)}}{J_z\sqrt{J_xJ_y}}\mathbf{Z} & \frac{\sqrt{\zeta(\zeta-J_y^2)}}{J_y\sqrt{J_zJ_x}}\mathbf{Y} \\
    0 & \frac{\sqrt{\zeta(\zeta-J_z^2)}}{J_z\sqrt{J_xJ_y}}\mathbf{Z} & \frac{\zeta}{J_y} & -\frac{\sqrt{\zeta(\zeta-J_x^2)}}{J_x\sqrt{J_yJ_z}}\mathbf{X} \\
    0 & -\frac{\sqrt{\zeta(\zeta-J_y^2)}}{J_y\sqrt{J_zJ_x}}\mathbf{Y} & \frac{\sqrt{\zeta(\zeta-J_x^2)}}{J_x\sqrt{J_yJ_z}}\mathbf{X} & \frac{\zeta}{J_z} \\
  \end{bmatrix}
  .
\end{equation}

\section{Methods}
\label{sec:methods}

Let me explain the methods that were used during the search. First, I comment on symmetry considerations that help to choose a form of $\mathcal{A}_j$, then I describe how to reduce the problem to a set of algebraic equations to solve.

\subsection{Symmetry considerations}
\label{subsec:symmetry}

Here I discuss how one could come up with to concrete form of MPO using symmetry considerations (although to be honest, one is more likely to adopt from \cite{Fendley_2025,Fukai_2026}).

Heisenberg spin chains include various symmetries, it is natural to search for integrals of motion that are also symmetric. First of all, there is a translational invariance that suggests the tensor $\mathcal{A}_j$ to be independent of the site $j$.

Translationally invariant operator matrices $\mathcal{A}$ has a \emph{gauge freedom} preserving MPO charge $\mathcal{M}$:
\begin{equation}
  \mathcal{A} \mapsto S \mathcal{A} S^{-1}, \qquad
  \adjustbox{raise=0.25em}{
  \begin{tikzpicture}
    \draw (-0.7,0) -- (0.7,0);
    \draw (0,-0.7) -- (0,0.7);
    \node[base,fill=blue!10] at (0,0) {$\mathcal{A}$};
  \end{tikzpicture}
  }\mapsto
  \adjustbox{raise=0.25em}{
  \begin{tikzpicture}
    \draw (-1.4,0) -- (1.5,0);
    \draw (0,-0.7) -- (0,0.7);
    \node [base] at (-0.8,0) {$S$} ;
    \node[base,fill=blue!10] at (0,0) {$\mathcal{A}$};
    \node [base] at (+0.9,0) {$S^{-1}$} ;
  \end{tikzpicture}
  }
  ,
\end{equation}
where $S$ is some complex matrix with scalar elements. Gauge can be fixed by symmetry requirements.

The XYZ Heisenberg model has $\Int_2\times\Int_2$ symmetry given by rotations $\mathbf{Z}^{\otimes L},\,\mathbf{Z}^{\otimes L},\,\mathbf{Z}^{\otimes L}$ acting on all spins. For MPO $\mathcal{M}$ to be invariant under such symmetries, I impose on $\mathcal{A}$ a condition that physical symmetries $U$ translate to gauge symmetries $S$:
\begin{equation}
\label{eq:physical_to_gauge}
  U^\dag \mathcal{A}_{\alpha,\beta} U = \sum_{\mu,\nu} S_{\alpha,\mu} \mathcal{A}_{\mu,\nu} (S^{-1})_{\nu,\beta},
  \qquad
  \adjustbox{raise=0.25em}{
  \begin{tikzpicture}
    \draw (-0.7,0) -- (0.7,0);
    \draw (0,-1.2) -- (0,1.2);
    \node [base] at (0,+0.7) {$U^\dag$} ;
    \node [base,fill=blue!10] at (0,0) {$\mathcal{A}$};
    \node [base] at (0,-0.7) {$U$} ;
  \end{tikzpicture}
  }
  \quad = \quad
  \adjustbox{raise=0.25em}{
  \begin{tikzpicture}
    \draw (-1.4,0) -- (1.5,0);
    \draw (0,-0.7) -- (0,0.7);
    \node [base] at (-0.8,0) {$S$} ;
    \node[base,fill=blue!10] at (0,0) {$\mathcal{A}$};
    \node [base] at (+0.9,0) {$S^{-1}$} ;
  \end{tikzpicture}
  }
  .
\end{equation}
That means, the group of symmetries acts as a representation on the space of local tensors $\mathcal{A}$. [For infinitesimal symmetries, Eq.~\eqref{eq:physical_to_gauge} easily translates to Lie algebra representations.] Choosing correct representation includes the choice of appropriate bond dimension. In the case I consider, the bond dimension is $4$ and the $\Int_2\times\Int_2$-representation is given by
\begin{equation}
\begin{aligned}
  & O \mapsto \mathbf{X} O \mathbf{X} \quad &\text{acts as }\quad
  &\mathcal{A}\mapsto
  \begin{bmatrix}1&&&\\ &1&&\\ &&-1&\\ &&&-1\end{bmatrix}
  \mathcal{A}
  \begin{bmatrix}1&&&\\ &1&&\\ &&-1&\\ &&&-1\end{bmatrix}
  ,
  \\
  & O \mapsto \mathbf{Y} O \mathbf{Y} \quad &\text{acts as }\quad
  &\mathcal{A}\mapsto
  \begin{bmatrix}1&&&\\ &-1&&\\ &&1&\\ &&&-1\end{bmatrix}
  \mathcal{A}
  \begin{bmatrix}1&&&\\ &-1&&\\ &&1&\\ &&&-1\end{bmatrix}
  ,
  \\
  & O \mapsto \mathbf{Z} O \mathbf{Z} \quad &\text{acts as }\quad
  &\mathcal{A}\mapsto
  \begin{bmatrix}1&&&\\ &-1&&\\ &&-1&\\ &&&1\end{bmatrix}
  \mathcal{A}
  \begin{bmatrix}1&&&\\ &-1&&\\ &&-1&\\ &&&1\end{bmatrix}
  .
\end{aligned}
\end{equation}
An operator matrix $\mathcal{A}$ satisfies this $\Int_2\times\Int_2$-symmetry if and only if it has form
\begin{equation}
  \mathcal{A} =
  \begin{bmatrix}
    v & q_x\,\mathbf{X} & q_y\,\mathbf{Y} & q_z\,\mathbf{Z} \\
    p_x\,\mathbf{X} & w_x & -r_{x,y}\,\mathbf{Z} & r_{x,z}\,\mathbf{Y} \\
    p_y\,\mathbf{Y} & r_{y,x}\,\mathbf{Z} & w_y & -r_{y,z}\,\mathbf{X} \\
    p_z\,\mathbf{Z} & -r_{z,x}\,\mathbf{Y} & r_{z,y}\,\mathbf{X} & w_z
  \end{bmatrix}
\end{equation}
where all the appeared elements $v,p_\bullet,q_\bullet,r_\bullet,w_\bullet$ (together with lower indices) are complex variables.

At this point, the gauge symmetry $S$ commutes with $\Int_2\times\Int_2$-symmetry if and only if it is a diagonal matrix $S=\mathrm{diag}(s_0,s_x,s_y,s_z)$. Additionally, the system exhibits an anti-unitary $\mathcal{PT}$-symmetry, which consists of the spatial mirroring of a spin chain (transposition of operator matrices $\mathcal{A}\mapsto\mathcal{A}^T$) together with time invertion (reflection of the spin directions $\mathbf{X}\mapsto -\mathbf{X}$, $\mathbf{Y}\mapsto -\mathbf{Y}$, $\mathbf{Z}\mapsto -\mathbf{Z}$). $\mathcal{PT}$-symmetry can be accounted by requiring that
\begin{equation}
  \begin{aligned}
    \mathbf{X}&\mapsto -\mathbf{X}, \\
    \mathbf{Y}&\mapsto -\mathbf{Y}, \\
    \mathbf{Z}&\mapsto -\mathbf{Z}
  \end{aligned}
  \qquad \text{acts as } \quad
  \mathcal{A}\mapsto
  \begin{bmatrix}1&&&\\ &-1&&\\ &&-1&\\ &&&-1\end{bmatrix}
  \mathcal{A}^T
  \begin{bmatrix}1&&&\\ &-1&&\\ &&-1&\\ &&&-1\end{bmatrix}
  ,
\end{equation}
which induces that lower $3\times3$ submatrix is anti-symmetric and $p_x=q_x$, $p_y=q_y$, $p_z=q_z$. Finally, the operator matrix $\mathcal{A}$ has form
\begin{equation}
\label{eq:A_parametrization}
  \mathcal{A} =
  \begin{bmatrix}
    v & p_x\,\mathbf{X} & p_y\,\mathbf{Y} & p_z\,\mathbf{Z} \\
    p_x\,\mathbf{X} & w_x & -r_z\,\mathbf{Z} & r_y\,\mathbf{Y} \\
    p_y\,\mathbf{Y} & r_z\,\mathbf{Z} & w_y & -r_x\,\mathbf{X} \\
    p_z\,\mathbf{Z} & -r_y\,\mathbf{Y} & r_x\,\mathbf{X} & w_z
  \end{bmatrix}
\end{equation}
and is parametrized by $10$ variables $\{v,w_x,w_y,w_z,p_x,p_y,p_z,r_x,r_y,r_z\}$. The error term $\mathcal{E}$ should be in the same form
\begin{equation}
\label{eq:E_parametrization}
  \mathcal{E}_j =
  \begin{bmatrix}
    \widetilde{v} & \widetilde{p}_x\,\mathbf{X}_j & \widetilde{p}_y\,\mathbf{Y}_j & \widetilde{p}_z\,\mathbf{Z}_j \\
    \widetilde{p}_x\,\mathbf{X}_j & \widetilde{w}_x & -\widetilde{r}_z\,\mathbf{Z}_j & \widetilde{r}_y\,\mathbf{Y}_j \\
    \widetilde{p}_y\,\mathbf{Y}_j & \widetilde{r}_z\,\mathbf{Z}_j & \widetilde{w}_y & -\widetilde{r}_x\,\mathbf{X}_j \\
    \widetilde{p}_z\,\mathbf{Z}_j & -\widetilde{r}_y\,\mathbf{Y}_j & \widetilde{r}_x\,\mathbf{X}_j & \widetilde{w}_z
  \end{bmatrix}
\end{equation}
and parametrized with $10$ variables $\{\widetilde{v},\widetilde{w}_x,\widetilde{w}_y,\widetilde{w}_z,\widetilde{p}_x,\widetilde{p}_y,\widetilde{p}_z,\widetilde{r}_x,\widetilde{r}_y,\widetilde{r}_z\}$.

The $\mathrm{U}(1)\ltimes\Int_2$-symmetry of XXZ model means that the tensor $\mathcal{A}$ has form
\begin{equation}
  \mathcal{A} =
  \begin{bmatrix}
    v & p_x\,\mathbf{X} & p_x\,\mathbf{Y} & p_z\,\mathbf{Z} \\
    p_x\,\mathbf{X} & w_x & -r_z\,\mathbf{Z} & r_x\,\mathbf{Y} \\
    p_x\,\mathbf{Y} & r_z\,\mathbf{Z} & w_x & -r_x\,\mathbf{X} \\
    p_z\,\mathbf{Z} & -r_x\,\mathbf{Y} & r_x\,\mathbf{X} & w_z
  \end{bmatrix}
\end{equation}
and depends on $7$ parameters $\{v,w_x,w_z,p_x,p_z,r_x,r_z\}$. Note that the solution Eq.~\eqref{eq:XXZ_plane_solution} does not follow this symmetry.

The $\mathrm{SU}(2)$-symmetry of XXX model means that the tensor $\mathcal{A}$ is written as
\begin{equation}
  \mathcal{A} =
  \begin{bmatrix}
    v & p\,\mathbf{X} & p\,\mathbf{Y} & p\,\mathbf{Z} \\
    p\,\mathbf{X} & w & -r\,\mathbf{Z} & r\,\mathbf{Y} \\
    p\,\mathbf{Y} & r\,\mathbf{Z} & w & -r\,\mathbf{X} \\
    p\,\mathbf{Z} & -r\,\mathbf{Y} & r\,\mathbf{X} & w
  \end{bmatrix}
\end{equation}
and depends on $4$ parameters $\{v,w,p,r\}$.

\subsection{System of equations}
\label{subsec:system_of_equations}

Having parametrized matrices $\mathcal{A}$ and $\mathcal{E}$ [see Eqs.~\eqref{eq:A_parametrization} and \eqref{eq:E_parametrization}], one can substitute $\mathcal{A}$ and $\mathcal{E}$ to main equation Eq.~\eqref{eq:main_equation}, obtaining a system of algebraic equations:
\begin{equation}
\label{eq:system_of_equations}
\begin{gathered}
  \begin{aligned}
   J_z p_z r_y+J_y p_y r_z&=v \tilde{p}_x-\tilde{v} p_x, &\quad&& J_y p_y p_z-J_z r_y r_z&=r_x \tilde{w}_y-w_y \tilde{r}_x, \\
   J_y p_z r_y+J_z p_y r_z&=w_x \tilde{p}_x-p_x \tilde{w}_x, &\quad&& J_z p_y p_z-J_y r_y r_z&=r_x \tilde{w}_z-w_z \tilde{r}_x, \\
   p_x \left(v J_z-J_y w_x\right)&=r_y \tilde{p}_z-p_z \tilde{r}_y, &\quad&& r_x \left(J_z w_y-J_y w_z\right)&=r_y \tilde{r}_z-r_z \tilde{r}_y, \\
   p_x \left(v J_y-J_z w_x\right)&=r_z \tilde{p}_y-p_y \tilde{r}_z, &\quad&& r_x \left(J_y w_y-J_z w_z\right)&=p_y \tilde{p}_z-p_z \tilde{p}_y, \\
  \end{aligned}
  \\[0.5em]
  \text{and cyclic shifts of $x,y,z$.}
\end{gathered}
\end{equation}
This is a set of $24$ homogeneous quadratic equations in $20$ variables. As shown in examples from Section~\ref{sec:results}, the system is degenerate and the solution set contains $4$-dimensional components (two meaningful parameters, one scaling parameter and one error term parameter $\gamma$). The variables of $\mathcal{E}$ fit linearly in the system and they can be eliminated, for example by setting $\widetilde{v} = 0$ (by the choice of $\gamma$) and using substitutions
\begin{equation}
\begin{gathered}
  \begin{aligned}
    \widetilde{w}_x &\to \frac{\widetilde{p}_x}{p_x} w_x - \frac{J_yr_yp_z+J_zr_zp_y}{p_x}, \\
    \widetilde{r}_x &\to \frac{\widetilde{p}_z}{p_z} r_x - \frac{J_xp_yw_y-v J_z p_y}{p_z}, \\
    \widetilde{p}_x &\to \frac{\widetilde{v}}{v} p_x + \frac{J_yr_yp_z+J_zr_zp_y}{v}, \\
  \end{aligned}
  \\[0.5em]
  \text{and cyclic shifts of $x,y,z$.}
\end{gathered}
\end{equation}
In principle, all solutions to system Eq.~\eqref{eq:system_of_equations} can be fully characterized using symbolic algebra and Gr{\"o}bner basis methods \cite{Cox_1991}. In practice, for XXX and XXZ cases the Gr{\"o}bner bases are quite manageable to obtain, but the XYZ case seems to require too much time and memory resources to be computed on a laptop. Also, approximate numerical solutions can be found using developed methods of semidefinite programming and rank minimization \cite{Dattorro_2019,Lemon_2016}.

When studying the Trotterized dynamics, substituting parametrizations \eqref{eq:A_parametrization} for operator matrices $\mathcal{A}_o$ and $\mathcal{A}_e$ to Eq.~\eqref{eq:trotterized_main_equation} also leads to a set of homogeneous quadratic equations, additionally dependent on Trotterization step $\tau$. These equations have much harder structure than in $\tau\to 0$ case, so it is profittable to first try solving Eq.~\eqref{eq:main_equation} and then to watch for generalization.

I found solutions described in Section~\ref{sec:results} with the help of computer algebra systems by sequentially choosing convenient changes of variables. For checking the correctness of the solutions and concrete computations consult Supplementary Materials \cite{Supplementary}.

\section{Discussion}
\label{sec:discussion}

As announced in the abstract, the scope of this work is limited to reporting on the found families of conserved quantities. Let me summarize the work and then discuss the questions left for future research and speculate about possible applications.

\subsection{Summary}
\label{subsec:summary}

Here is a brief summary of the work. I wrote down solutions to Eq.~\eqref{eq:main_equation} for Heisenberg spin chains with periodic boundary conditions and various kinds of anisotropy. When searching for these families of charges, I chose parametrizations of MPOs using ideas of symmetry, wrote down a set of algebraic equations on the parameters and solved them using computer algebra.

MPO charges for XXX and XXZ models have natural parametrizations over points of a sphere. Integrals of motions for XXX, XXZ, XX and XY models are proven to be stable under Trotterizations, because they can be generalized to solve Eq.~\eqref{eq:trotterized_main_equation}. Most importantly, I was able to find a family of MPO conserved quantities for XYZ model [Eq.~\eqref{eq:XYZ_solution}]. Other solutions can be obtained from it as limiting cases (though, one should be careful about correctly choosing the limits).

\subsection{Outlook}
\label{subsec:outlook}

First of all, it is interesting to check if the listed solutions completely describe all local integrals of motion. One possible proof direction might be to use the following combinatorial idea. MPO operators can be generated by finite-state automata (oriented graphs with labeled edges) \cite{Crosswhite_2008}, each term in the representation of MPO as a sum of Pauli strings corresponds to a path on the automaton (for periodic boundary conditions, a cycle). In the case considered in this paper, the vertices of a graph are $\{0,x,y,z\}$, and jumping from one vertex to another corresponds to multiplying a string by $\mathbf{X},\mathbf{Y},\mathbf{Z}$ together with multiplication by scalars. Collecting paths corresponding to different weights, one might try to find a combinatorial characterization of local charges and check the known recursive relations \cite{Grabowski_1994,Yamada_2023}. That study might as well give new insights on combinatorial identities between local charges \cite{Grabowski_1994,Nozawa_2020,Yamada_2023,Fukai_2026}.

Also, one might try to check completeness and give explicit characterization to local Floquet solutions of Heisenberg spin chains using MPO families found it this work.

For Heisenberg spin chains, there is a developed theory of quasi-local charges \cite{Ilievski_2016,Prosen_2013,Prosen_2014,Ilievski_2015,Zadnik_2016}, which carry more information about the system than local charges. The information they carry is roughly equivalent to Theormodynamic Bethe ansatz. It would be surprising if the solutions found in this work are also complete in the class of quasi-local charges, but probably they are not.

Suppose that the completeness of some family of integrals of motion $\mathcal{M} = \mathcal{M}(\theta)$ dependent on parameter $\theta$ is proven. This result can help in developing new methods for studying dynamics of integrable models. Given a quantum state $\rho$, the set of average values $\langle \mathcal{M}(\theta) \rangle_{\rho}$ is conserved during evolution and might completely characterize state's trajectory in time. It would then be possible to study dynamical properties of the model it terms of function $\theta\mapsto\langle \mathcal{M}(\theta) \rangle_{\rho}$ defined on the parameter space (for example, a sphere). If each $\mathcal{M}(\theta)$ is in MPO form, this function can be efficiently computed for product states or for matrix product states, sometimes also in the thermodynamic limit.

More generally, it might be possible to use the approach of generalized Gibbs ensembles \cite{Vidmar_2016,Pozsgay_2013,Costeniuc_2005_1,Costeniuc_2006_2} in connection with MPO charges for applications to integrable models. In this approach, the known integrals of motion of the system are included into the definition of thermodynamic ensemble, allowing for describing equilibrium states using generalized temperatures. Provided that one has an analytical characterization of all charges in an integrable model, one can describe many properties of the system using generalized temperatures.

Talking about XYZ family of charges Eq.~\eqref{eq:XYZ_solution}, it might be useful to further explore its structure, for example to search for new natural parametrizations (possibly using elliptic functions), try to find parametrization depending on points of a sphere, and to check if there are some interesting limits that were not mentioned in this work. Also, it is important to find Floquet integrals of motion generalizing this solution. I conjecture that even matrices $\mathcal{A}_o$ is can be found from $\mathcal{A}$ by deforming anisotropies
\begin{equation}
  J_x^\tau = \frac{\sin(J_x\tau)}{\sin\tau}, \qquad
  J_y^\tau = \frac{\sin(J_y\tau)}{\sin\tau}, \qquad
  J_z^\tau = \frac{\sin(J_z\tau)}{\sin\tau},
\end{equation}
but finding the corresponding $\mathcal{A}_e$ is not that simple.

It would be interesting to check if there are other useful families of MPO integrals of motions in Heisenberg models. In the XXX case, it is possible to list all bond dimension $4$ solutions and find that there are no other interesting ones except Eq.~\eqref{eq:XXX_solution}.

Certainly, the methods described in this work can be applied to more general models of many-body physics: most importantly, to describe integrals of motion of $\mathrm{SO}(4)$-invariant one-dimensional Hubbard model \cite{Essler_2005,Fukai_2023,Fukai_2024_1,Fukai_2024_2}. The authors of \cite{Fukai_2026} announced that they are working in this direction.

The idea of studying tensor network observables introduced in \cite{Fendley_2025} can have very broad setting. In its full generality, the method can be described as consisting of the following steps:
\begin{enumerate}[label=\textbf{Step~\arabic*.},start=0,itemindent=2\parindent,leftmargin=\parindent,rightmargin=\parindent]
  \item Choose a physical system with defined dynamical process. Look over its global geometry and symmetries. (For example, spin-$\frac{1}{2}$ chains with periodic boundary conditions and dynamics given by Heisenberg Hamiltonian.)
  \item Choose a tensor network architecture for observables in accordance with global geometry of the system (for $1$-dimensional systems choose MPO, for $2$-dimensional square lattices choose PEPO, and so on).  Write down the equations on local tensors that ensure the observables preserve their structure under dynamics [such as Eqs.~\eqref{eq:main_equation} and \eqref{eq:trotterized_main_equation}].
  \item Choose a parametrization for the observables, this parametrization should reprect symmetries of the system [such as Eq.~\eqref{eq:A_parametrization}]. Write down the system equations on the parameters [see Eq.~\eqref{eq:system_of_equations}].
  \item Solve the resulting equations.
\end{enumerate}
The last step is, of course, crucial. If the set of equations contains many variables and constraints, solving it is practically impossible. Still, if the physical system is symmetric enough, then after considerable effort it might be solved.

This method seems to be applicable at least to one-dimensional spin chains with nearest-neighbour interactions with some form of translation invariance, with general boundary conditions, and with some defects. It would be nice to find applications to more sophisticated systems, for example some $2$-dimensional exactly solvable model such as Kitaev honeycomb model \cite{Kitaev_2006}. Also, it is interesting to apply the theory of fermionic tensor networks \cite{Mortier_2025} for such methods.

The approach can be used to study not only the conserved quantities, but also some dynamically evolving observables. For example, the Eq.~\eqref{eq:main_equation} can be generalized to express the exact dynamics of observables preserving MPO form during the evolution. Suppose that an MPO $\mathcal{M}$ is defined by a sequence of operator matrices $\mathcal{A}_j$ over sites $j$, and suppose that there exist operator matrices $\mathcal{L}_{j}$ and $\mathcal{R}_{j+1}$ such that
\begin{equation}
\label{eq:generalized_main_equation}
  i \;
  \adjustbox{raise=0.25em}{
  \begin{tikzpicture}
    \draw (-0.7,0) -- (1.7,0);
    \draw (0,-0.7) -- (0,1.6); \draw (1,-0.7) -- (1,1.6);
    \draw[base] (-0.3,0.6) rectangle (1.3,1.3) node[midway] {$H_{j,j+1}$};
    \node[base,fill=blue!10] at (0,0) {$\mathcal{A}_j$};
    \node[base,fill=blue!10] at (1,0) {$\mathcal{A}_{j\!+\!1}$};
  \end{tikzpicture}
  }
  \;
  -
  i \;
  \adjustbox{raise=0.25em}{
  \begin{tikzpicture}
    \draw (-0.7,0) -- (1.7,0);
    \draw (0,0.7) -- (0,-1.6); \draw (1,0.7) -- (1,-1.6);
    \node[base,fill=blue!10] at (0,0) {$\mathcal{A}_j$};
    \node[base,fill=blue!10] at (1,0) {$\mathcal{A}_{j\!+\!1}$};
    \draw[base] (-0.3,-0.6) rectangle (1.3,-1.3) node[midway] {$H_{j,j+1}$};
  \end{tikzpicture}
  }
  \quad = \quad
  \adjustbox{raise=0.25em}{
  \begin{tikzpicture}
    \draw (-0.7,0) -- (1.7,0);
    \draw (0,-0.7) -- (0,0.7); \draw (1,-0.7) -- (1,0.7);
    \node[base,fill=red!10] at (0,0) {$\mathcal{L}_j$} ;
    \node[base,fill=blue!10] at (1,0) {$\mathcal{A}_{j\!+\!1}$};
  \end{tikzpicture}
  }
  -
  \adjustbox{raise=0.25em}{
  \begin{tikzpicture}
    \draw (-0.7,0) -- (1.7,0);
    \draw (0,-0.7) -- (0,0.7); \draw (1,-0.7) -- (1,0.7);
    \node[base,fill=blue!10] at (0,0) {$\mathcal{A}_j$};
    \node [base, fill=green!10] at (1,0) {$\mathcal{R}_{j+1}$} ;
  \end{tikzpicture}
  }
  .
\end{equation}
In this case, Heisenberg equation guarantees that the MPO form of $\mathcal{M} = \mathcal{M}(t)$ is preserved during the evolution (at least for small times $t$) and $\frac{d}{dt}\mathcal{A}_j = \mathcal{L}_j-\mathcal{R}_j$. It should be natural to search for time-evolving MPO observables of the XX-model \cite{Lychkovskiy_2021,Teretenkov_2024}. Preliminary investigation indicates that for dynamically evolving MPO observables to exist, they should somehow break the symmetry of the integrable model.

The approach of studying MPO conserved quantities was inspired by the search for strong zero modes in integrable systems \cite{Fendley_2016,Essler_2025,Gehrmann_2026,Klobas_2023}, including Floquet systems \cite{Vernier_2024}. Thus, establishing the theory of MPO integrals of motion may lead to new results on strong zero modes, which are interesting by themselves and have applications to topological codes in quantum error correction \cite{Sarma_2015}.

In Section~\ref{subsec:system_of_equations}, I discussed that Eq.~\eqref{eq:main_equation} reduces to a set of quadratic algebraic equations, and such equations could be solved numerically using semidefinite optimization. I assume that the development of numerical methods for the search of approximate integrals of motion might be an important direction on its own. It might be applicable for theoretical and experimental studies in many-body systems, for searching indications of integrability versus chaos, and strong zero modes. It might also have importance to Hidden Subgroup problems \cite{Hinsche_2025}.

Finally, it is possible to search for applications in quantum computing, especially for studying structural properties of brick-wall circuits and variational ansatzes \cite{Fisher_2023,Vasseur_2026}. Note that the well-studied class of dual unitary circuits includes local XXZ interactions \cite{Bertini_2019,Piroli_2020}, and $U(1)$-invariant two-step Floquet protocols were recently found to be integrable \cite{Znidaric_2025}, making solutions Eqs.~\eqref{eq:XXZ_solution} and \eqref{eq:XXZ_floquet_solution} somewhat more relevant.

\section*{Acknowledgements}
I am grateful to Oleg V.~Lychkovskiy for his interest in the topic and for many useful discussions, as well as to Denis V.~Kurlov for introducing me to symbolic algebra methods in many-body physics. I thank L.~Zadnik for his comments on history and current status of integrable quantum circuits and the theory of quasi-local charges.


\paragraph{Funding information}
The work was supported by the Foundation for the Advancement of Theoretical Physics and Mathematics ``BASIS''. 



\bibliography{bibliography.bib}

@article{Baxter_1973_1,
  title={Eight-vertex model in lattice statistics and one-dimensional anisotropic {H}eisenberg chain. {I.} Some fundamental eigenvectors},
  journal={Annals of Physics},
  volume={76},
  number={1},
  pages={1-24},
  year={1973},
  issn={0003-4916},
  doi={10.1016/0003-4916(73)90439-9},
  author={Rodney Baxter}
}

@article{Baxter_1973_2,
  title={Eight-vertex model in lattice statistics and one-dimensional anisotropic {H}eisenberg chain. {II.} Equivalence to a generalized ice-type lattice model},
  journal={Annals of Physics},
  volume={76},
  number={1},
  pages={25-47},
  year={1973},
  issn={0003-4916},
  doi={10.1016/0003-4916(73)90440-5},
  author={Rodney Baxter}
}

@article{Baxter_1973_3,
  title={Eight-vertex model in lattice statistics and one-dimensional anisotropic {H}eisenberg chain. {III.} Eigenvectors of the transfer matrix and hamiltonian},
  journal={Annals of Physics},
  volume={76},
  number={1},
  pages={48-71},
  year={1973},
  issn={0003-4916},
  doi={10.1016/0003-4916(73)90441-7},
  author={Rodney Baxter},
}

@book{Baxter_2007,
  title={Exactly Solved Models in Statistical Mechanics},
  author={Baxter, R.J.},
  isbn={9780486462714},
  lccn={2007037510},
  series={Dover books on physics},
  year={2007},
  publisher={Dover Publications},
  url={https://store.doverpublications.com/products/9780486462714}
}

@article{Bertini_2019,
  title={Exact Correlation Functions for Dual-Unitary Lattice Models in $1+1$ Dimensions},
  volume={123},
  ISSN={1079-7114},
  url={http://dx.doi.org/10.1103/PhysRevLett.123.210601},
  DOI={10.1103/physrevlett.123.210601},
  number={21},
  journal={Physical Review Letters},
  publisher={American Physical Society (APS)},
  author={Bertini, Bruno and Kos, Pavel and Prosen, Tomaž},
  year={2019},
  month=nov
}

@article{Bethe_1931,
  author={Bethe, Hans},
  title={On the theory of metals. 1. Eigenvalues and eigenfunctions for
  the linear atomic chain},
  journal={Zeitschrift für Physik},
  volume={71},
  pages={205--226},
  year={1931},
  doi={10.1007/BF01341708}
}

@misc{Biamonte_2020,
  title={Lectures on Quantum Tensor Networks},
  author={Jacob Biamonte},
  year={2020},
  eprint={1912.10049},
  archivePrefix={arXiv},
  primaryClass={quant-ph},
  url={https://arxiv.org/abs/1912.10049},
}

@article{Costeniuc_2005_1,
  title={The Generalized Canonical Ensemble and Its Universal Equivalence with the Microcanonical Ensemble},
  volume={119},
  ISSN={1572-9613},
  url={http://dx.doi.org/10.1007/s10955-005-4407-0},
  DOI={10.1007/s10955-005-4407-0},
  number={5–6},
  journal={Journal of Statistical Physics},
  publisher={Springer Science and Business Media LLC},
  author={Costeniuc, Marius and Ellis, Richard S. and Touchette, Hugo and Turkington, Bruce},
  year={2005},
  month=jun,
  pages={1283–1329}
}

@article{Costeniuc_2006_2,
  title={Generalized canonical ensembles and ensemble equivalence},
  volume={73},
  ISSN={1550-2376},
  url={http://dx.doi.org/10.1103/PhysRevE.73.026105},
  DOI={10.1103/physreve.73.026105},
  number={2},
  journal={Physical Review E},
  publisher={American Physical Society (APS)},
  author={Costeniuc, M. and Ellis, R. S. and Touchette, H. and Turkington, B.},
  year={2006},
  month=feb
}

@book{Cox_1991,
  title={Ideals, Varieties, and Algorithms: An Introduction to Computational Algebraic Geometry and Commutative Algebra},
  author={Cox, David A. and Little, John and O'Shea Donal},
  series={Undergraduate Texts in Mathematics},
  doi={10.1007/978-3-031-91841-4},
  publisher={Springer Cham},
  isbn={978-3-031-91840-7},
  issn={0172-6056},
  edition={5},
  year={2025}
}

@article{Crosswhite_2008,
  title={Finite automata for caching in matrix product algorithms},
  author={Crosswhite, Gregory M. and Bacon, Dave},
  journal={Phys. Rev. A},
  volume={78},
  issue={1},
  pages={012356},
  numpages={13},
  year={2008},
  month={Jul},
  publisher={American Physical Society},
  doi={10.1103/PhysRevA.78.012356},
  url={https://link.aps.org/doi/10.1103/PhysRevA.78.012356}
}

@misc{Cuiper_2026,
  title={{L}es {H}ouches Lecture Notes on Tensor Networks},
  author={Bram Vancraeynest-De Cuiper and Weronika Wiesiolek and Frank Verstraete},
  year={2026},
  eprint={2512.24390},
  archivePrefix={arXiv},
  primaryClass={cond-mat.str-el},
  url={https://arxiv.org/abs/2512.24390}
}

@book{Dattorro_2019,
  title={Convex Optimization \& Euclidean Distance Geometry},
  author={Dattorro, Jon},
  year={2019},
  publisher={Meboo Publishing USA},
  note={Version 2019.10.28},
  url={https://meboo.convexoptimization.com/access.html}
}

@article{Destri_1987,
  title={Light-cone lattice approach to fermionic theories in {2D}: The massive Thirring model},
  journal={Nuclear Physics B},
  volume={290},
  pages={363-391},
  year={1987},
  issn={0550-3213},
  doi={10.1016/0550-3213(87)90193-3},
  url={https://www.sciencedirect.com/science/article/pii/0550321387901933},
  author={Destri, C. and De Vega, H. J.}
}

@book{Essler_2005,
  title={The one-dimensional {H}ubbard model},
  author={Essler, Fabian H L and Frahm, Holger and G{\"o}hmann, Frank and Kl{\"u}mper, Andreas and Korepin, Vladimir E},
  year={2005},
  publisher={Cambridge University Press},
  doi={10.1017/CBO9780511534843}
}

@misc{Essler_2025,
  title={Strong zero modes in integrable spin-{S} chains},
  author={Fabian H. L. Essler and Paul Fendley and Eric Vernier},
  year={2025},
  eprint={2512.07742},
  archivePrefix={arXiv},
  primaryClass={cond-mat.stat-mech},
  url={https://arxiv.org/abs/2512.07742},
}

@article{Faddeev_1994,
  author={Faddeev, L. and Volkov, A. Yu.},
  year={1994},
  month=oct,
  title={{H}irota equation as an example of an integrable symplectic map},
  journal={Latters in Mathematical Physics},
  volume={32},
  pages={125--135},
  issn={1573-0530},
  url={https://doi.org/10.1007/BF00739422},
  doi={10.1007/BF00739422}
}

@article{Fendley_2016,
  title={Strong zero modes and eigenstate phase transitions in the {XYZ}/interacting {M}ajorana chain},
  volume={49},
  ISSN={1751-8121},
  url={http://dx.doi.org/10.1088/1751-8113/49/30/30LT01},
  DOI={10.1088/1751-8113/49/30/30lt01},
  number={30},
  journal={Journal of Physics A: Mathematical and Theoretical},
  publisher={IOP Publishing},
  author={Fendley, Paul},
  year={2016},
  month=jun, pages={30LT01}
}

@misc{Fendley_2025,
  title={{XYZ} integrability the easy way},
  author={Paul Fendley and Sascha Gehrmann and Eric Vernier and Frank Verstraete},
  year={2025},
  eprint={2511.04674},
  archivePrefix={arXiv},
  primaryClass={cond-mat.stat-mech},
  url={https://arxiv.org/abs/2511.04674},
}

@article{Fisher_2023,
  title={Random Quantum Circuits},
  volume={14},
  ISSN={1947-5462},
  url={http://dx.doi.org/10.1146/annurev-conmatphys-031720-030658},
  DOI={10.1146/annurev-conmatphys-031720-030658},
  number={1},
  journal={Annual Review of Condensed Matter Physics},
  publisher={Annual Reviews},
  author={Fisher, Matthew P.A. and Khemani, Vedika and Nahum, Adam and Vijay, Sagar},
  year={2023},
  month=mar,
  pages={335–379}
}

@article{Fukai_2023,
  title={All Local Conserved Quantities of the One-Dimensional {H}ubbard Model},
  volume={131},
  ISSN={1079-7114},
  url={http://dx.doi.org/10.1103/PhysRevLett.131.256704},
  DOI={10.1103/physrevlett.131.256704},
  number={25},
  journal={Physical Review Letters},
  publisher={American Physical Society (APS)},
  author={Fukai, Kohei},
  year={2023},
  month=dec
}

@article{Fukai_2024_1,
  title={Proof of Completeness of the Local Conserved Quantities in the One-Dimensional {H}ubbard Model},
  volume={191},
  ISSN={1572-9613},
  url={http://dx.doi.org/10.1007/s10955-024-03267-y},
  DOI={10.1007/s10955-024-03267-y},
  number={6},
  journal={Journal of Statistical Physics},
  publisher={Springer Science and Business Media LLC},
  author={Fukai, Kohei},
  year={2024},
  month=jun
}

@misc{Fukai_2024_2,
  title={Study of local conserved quantities in the one-dimensional {H}ubbard model},
  author={Kohei Fukai},
  year={2024},
  eprint={2402.08924},
  archivePrefix={arXiv},
  primaryClass={cond-mat.stat-mech},
  url={https://arxiv.org/abs/2402.08924},
}

@misc{Fukai_2026,
  title={Matrix product operator representations for the local conserved quantities of the spin-$1/2$ {XYZ} chain},
  author={Kohei Fukai and Kyoichi Yamada},
  year={2026},
  eprint={2601.09245},
  archivePrefix={arXiv},
  primaryClass={nlin.SI},
  url={https://arxiv.org/abs/2601.09245},
}

@misc{Gehrmann_2026,
  title={Exact strong zero modes in quantum circuits and spin chains with non-diagonal boundary conditions},
  author={Sascha Gehrmann and Fabian H. L. Essler},
  year={2026},
  eprint={2511.05490},
  archivePrefix={arXiv},
  primaryClass={cond-mat.stat-mech},
  url={https://arxiv.org/abs/2511.05490},
}

@article{Grabowski_1994,
  title={QUANTUM INTEGRALS OF MOTION FOR THE {H}EISENBERG SPIN CHAIN},
  volume={09},
  ISSN={1793-6632},
  url={http://dx.doi.org/10.1142/S0217732394002057},
  DOI={10.1142/s0217732394002057},
  number={24},
  journal={Modern Physics Letters A},
  publisher={World Scientific Pub Co Pte Lt},
  author={Grabowski, Marek P. and Mathieu, Pierre},
  year={1994},
  month=aug, pages={2197–2206}
}

@article{Grabowski_1995,
  title={Structure of the Conservation Laws in Quantum Integrable Spin Chains with Short Range Interactions},
  volume={243},
  ISSN={0003-4916},
  url={http://dx.doi.org/10.1006/aphy.1995.1101},
  DOI={10.1006/aphy.1995.1101},
  number={2},
  journal={Annals of Physics},
  publisher={Elsevier BV},
  author={Grabowski, M.P. and Mathieu, P.},
  year={1995},
  month=nov,
  pages={299–371}
}

@article{Gritsev_2017,
  title={Integrable {F}loquet dynamics},
  volume={2},
  ISSN={2542-4653},
  url={http://dx.doi.org/10.21468/SciPostPhys.2.3.021},
  DOI={10.21468/scipostphys.2.3.021},
  number={3},
  journal={SciPost Physics},
  publisher={Stichting SciPost},
  author={Gritsev, Vladimir and Polkovnikov, Anatoli},
  year={2017},
  month=jun
}

@article{Haldar_2021,
  title={Dynamical Freezing and Scar Points in Strongly Driven {F}loquet Matter: Resonance vs Emergent Conservation Laws},
  author={Haldar, Asmi and Sen, Diptiman and Moessner, Roderich and Das, Arnab},
  journal={Phys. Rev. X},
  volume={11},
  issue={2},
  pages={021008},
  numpages={25},
  year={2021},
  month={Apr},
  publisher={American Physical Society},
  doi={10.1103/PhysRevX.11.021008},
  url={https://link.aps.org/doi/10.1103/PhysRevX.11.021008}
}

@article{Heisenberg_1928,
  title={Zur theorie des ferromagnetismus},
  author={Heisenberg, Werner},
  journal={Zeitschrift f{\"u}r Physik},
  volume={49},
  number={9},
  pages={619--636},
  year={1928},
  publisher={Springer},
  doi={10.1007/BF01328601}
}

@misc{Hinsche_2025,
  title={The abelian state hidden subgroup problem: Learning stabilizer groups and beyond},
  author={Marcel Hinsche and Jens Eisert and Jose Carrasco},
  year={2025},
  eprint={2505.15770},
  archivePrefix={arXiv},
  primaryClass={quant-ph},
  url={https://arxiv.org/abs/2505.15770},
}

@article{Ilievski_2015,
  title={Quasilocal Conserved Operators in the Isotropic {H}eisenberg Spin-{$1/2$} Chain},
  volume={115},
  ISSN={1079-7114},
  url={http://dx.doi.org/10.1103/PhysRevLett.115.120601},
  DOI={10.1103/physrevlett.115.120601},
  number={12},
  journal={Physical Review Letters},
  publisher={American Physical Society (APS)},
  author={Ilievski, Enej and Medenjak, Marko and Prosen, Tomaž},
  year={2015},
  month=sep
}

@article{Ilievski_2016,
  title={Quasilocal charges in integrable lattice systems},
  volume={2016},
  ISSN={1742-5468},
  url={http://dx.doi.org/10.1088/1742-5468/2016/06/064008},
  DOI={10.1088/1742-5468/2016/06/064008},
  number={6},
  journal={Journal of Statistical Mechanics: Theory and Experiment},
  publisher={IOP Publishing},
  author={Ilievski, Enej and Medenjak, Marko and Prosen, Tomaž and Zadnik, Lenart},
  year={2016},
  month={jun},
  pages={064008}
}

@article{Katsura_2015,
  title={On integrable matrix product operators with bond dimension {$D=4$}},
  volume={2015},
  ISSN={1742-5468},
  url={http://dx.doi.org/10.1088/1742-5468/2015/01/P01006},
  DOI={10.1088/1742-5468/2015/01/p01006},
  number={1},
  journal={Journal of Statistical Mechanics: Theory and Experiment},
  publisher={IOP Publishing},
  author={Katsura, Hosho},
  year={2015},
  month=jan,
  pages={P01006}
}

@article{Kitaev_2006,
  title={Anyons in an exactly solved model and beyond},
  volume={321},
  ISSN={0003-4916},
  url={http://dx.doi.org/10.1016/j.aop.2005.10.005},
  DOI={10.1016/j.aop.2005.10.005},
  number={1},
  journal={Annals of Physics},
  publisher={Elsevier BV},
  author={Kitaev, Alexei},
  year={2006},
  month=jan,
  pages={2–111}
}

@article{Klobas_2023,
  title={Stochastic strong zero modes and their dynamical manifestations},
  volume={107},
  ISSN={2470-0053},
  url={http://dx.doi.org/10.1103/PhysRevE.107.L042104},
  DOI={10.1103/physreve.107.l042104},
  number={4},
  journal={Physical Review E},
  publisher={American Physical Society (APS)},
  author={Klobas, Katja and Fendley, Paul and Garrahan, Juan P.},
  year={2023},
  month=apr
}

@article{Lemon_2016,
  author={Lemon, Alex and So, Anthony Man-Cho and Ye, Yinyu},
  title={Low-Rank Semidefinite Programming: Theory and Applications},
  year={2016},
  issue_date={Aug 2016},
  publisher={Now Publishers Inc.},
  address={Hanover, MA, USA},
  volume={2},
  number={1–2},
  issn={2167-3888},
  url={https://doi.org/10.1561/2400000009},
  doi={10.1561/2400000009},
  journal={Found. Trends Optim.},
  month=aug,
  pages={1–156},
  numpages={160}
}

@article{Ljubotina_2019,
  title={Ballistic Spin Transport in a Periodically Driven Integrable Quantum System},
  volume={122},
  ISSN={1079-7114},
  url={http://dx.doi.org/10.1103/PhysRevLett.122.150605},
  DOI={10.1103/physrevlett.122.150605},
  number={15},
  journal={Physical Review Letters},
  publisher={American Physical Society (APS)},
  author={Ljubotina, Marko and Zadnik, Lenart and Prosen, Tomaž},
  year={2019},
  month=apr
}

@article{Lotkov_2022,
  title={{F}loquet integrability and long-range entanglement generation in the one-dimensional quantum {P}otts model},
  volume={105},
  ISSN={2469-9969},
  url={http://dx.doi.org/10.1103/PhysRevB.105.144306},
  DOI={10.1103/physrevb.105.144306},
  number={14},
  journal={Physical Review B},
  publisher={American Physical Society (APS)},
  author={Lotkov, A. I. and Gritsev, V. and Fedorov, A. K. and Kurlov, D. V.},
  year={2022},
  month=apr
}

@misc{Lu_2025,
  title={Dynamical freezing and enhanced magnetometry in an interacting spin ensemble},
  author={Ya-Nan Lu and Dong Yuan and Yixuan Ma and Yan-Qing Liu and Si Jiang and Xiang-Qian Meng and Yi-Jie Xu and Xiu-Ying Chang and Chong Zu and Hong-Zheng Zhao and Dong-Ling Deng and Lu-Ming Duan and Pan-Yu Hou},
  year={2025},
  eprint={2507.22982},
  archivePrefix={arXiv},
  primaryClass={quant-ph},
  url={https://arxiv.org/abs/2507.22982},
}

@article{Lychkovskiy_2021,
  title={Closed hierarchy of {H}eisenberg equations in integrable models with {O}nsager algebra},
  volume={10},
  ISSN={2542-4653},
  url={http://dx.doi.org/10.21468/SciPostPhys.10.6.124},
  DOI={10.21468/scipostphys.10.6.124},
  number={6},
  journal={SciPost Physics},
  publisher={Stichting SciPost},
  author={Lychkovskiy, Oleg},
  year={2021},
  month=jun
}

@article{Miao_2023,
  title={Integrable Quantum Circuits from the Star-Triangle Relation},
  volume={7},
  ISSN={2521-327X},
  url={http://dx.doi.org/10.22331/q-2023-11-03-1160},
  DOI={10.22331/q-2023-11-03-1160},
  journal={Quantum},
  publisher={Verein zur Forderung des Open Access Publizierens in den Quantenwissenschaften},
  author={Miao, Yuan and Vernier, Eric},
  year={2023},
  month=nov,
  pages={1160}
}

@article{Miao_2024,
  title={The {F}loquet {B}axterisation},
  volume={16},
  ISSN={2542-4653},
  url={http://dx.doi.org/10.21468/SciPostPhys.16.3.078},
  DOI={10.21468/scipostphys.16.3.078},
  number={3},
  journal={SciPost Physics},
  publisher={Stichting SciPost},
  author={Miao, Yuan and Gritsev, Vladimir and Kurlov, Denis V.},
  year={2024},
  month=mar
}

@article{Mortier_2025,
  title={Fermionic tensor network methods},
  volume={18},
  ISSN={2542-4653},
  url={http://dx.doi.org/10.21468/SciPostPhys.18.1.012},
  DOI={10.21468/scipostphys.18.1.012},
  number={1},
  journal={SciPost Physics},
  publisher={Stichting SciPost},
  author={Mortier, Quinten and Devos, Lukas and Burgelman, Lander and Vanhecke, Bram and Bultinck, Nick and Verstraete, Frank and Haegeman, Jutho and Vanderstraeten, Laurens},
  year={2025},
  month=jan
}

@article{Mukherjee_2026,
  title={{F}loquet Thermalization via Instantons near Dynamical Freezing},
  author={Mukherjee, Rohit and Guo, Haoyu and Chowdhury, Debanjan},
  journal={Phys. Rev. X},
  volume={16},
  issue={1},
  pages={011041},
  numpages={26},
  year={2026},
  month={Feb},
  publisher={American Physical Society},
  doi={10.1103/4w5w-57my},
  url={https://link.aps.org/doi/10.1103/4w5w-57my}
}

@article{Nozawa_2020,
  title={Explicit Construction of Local Conserved Quantities in the {XYZ} Spin-{$1/2$} Chain},
  volume={125},
  ISSN={1079-7114},
  url={http://dx.doi.org/10.1103/PhysRevLett.125.090602},
  DOI={10.1103/physrevlett.125.090602},
  number={9},
  journal={Physical Review Letters},
  publisher={American Physical Society (APS)},
  author={Nozawa, Yuji and Fukai, Kouhei},
  year={2020},
  month=aug
}

@article{Nienhuis_2021,
  title={The local conserved quantities of the closed {XXZ} chain},
  volume={54},
  ISSN={1751-8121},
  url={http://dx.doi.org/10.1088/1751-8121/ac0961},
  DOI={10.1088/1751-8121/ac0961},
  number={30},
  journal={Journal of Physics A: Mathematical and Theoretical},
  publisher={IOP Publishing},
  author={Nienhuis, Bernard and Huijgen, Onno E},
  year={2021},
  month=jun,
  pages={304001}
}

@article{Orbach_1958,
  title={Linear Antiferromagnetic Chain with Anisotropic Coupling},
  author={Orbach, R.},
  journal={Phys. Rev.},
  volume={112},
  issue={2},
  pages={309--316},
  numpages={0},
  year={1958},
  month={Oct},
  publisher={American Physical Society},
  doi={10.1103/PhysRev.112.309},
  url={https://link.aps.org/doi/10.1103/PhysRev.112.309}
}

@article{Orus_2014,
  title={A practical introduction to tensor networks: Matrix product states and projected entangled pair states},
  journal={Annals of Physics},
  volume={349},
  pages={117-158},
  year={2014},
  issn={0003-4916},
  doi={https://doi.org/10.1016/j.aop.2014.06.013},
  url={https://www.sciencedirect.com/science/article/pii/S0003491614001596},
  author={Román Orús}
}

@article{Paletta_2025,
  title={Integrability and charge transport in asymmetric quantum-circuit geometries},
  volume={58},
  ISSN={1751-8121},
  url={http://dx.doi.org/10.1088/1751-8121/ade483},
  DOI={10.1088/1751-8121/ade483},
  number={27},
  journal={Journal of Physics A: Mathematical and Theoretical},
  publisher={IOP Publishing},
  author={Paletta, Chiara and Duh, Urban and Pozsgay, Balázs and Zadnik, Lenart},
  year={2025},
  month=jul,
  pages={275001}
}

@misc{Perk_2017,
  title={{O}nsager algebra and cluster {XY}-models in a transverse magnetic field},
  author={Jacques H. H. Perk},
  year={2017},
  eprint={1710.03384},
  archivePrefix={arXiv},
  primaryClass={cond-mat.stat-mech},
  url={https://arxiv.org/abs/1710.03384},
}

@article{Piroli_2020,
  title={Exact dynamics in dual-unitary quantum circuits},
  volume={101},
  ISSN={2469-9969},
  url={http://dx.doi.org/10.1103/PhysRevB.101.094304},
  DOI={10.1103/physrevb.101.094304},
  number={9},
  journal={Physical Review B},
  publisher={American Physical Society (APS)},
  author={Piroli, Lorenzo and Bertini, Bruno and Cirac, J. Ignacio and Prosen, Tomaž},
  year={2020},
  month=mar
}

@article{Pirvu_2010,
  title={Matrix product operator representations},
  volume={12},
  ISSN={1367-2630},
  url={http://dx.doi.org/10.1088/1367-2630/12/2/025012},
  DOI={10.1088/1367-2630/12/2/025012},
  number={2},
  journal={New Journal of Physics},
  publisher={IOP Publishing},
  author={Pirvu, B and Murg, V and Cirac, J I and Verstraete, F},
  year={2010},
  month=feb,
  pages={025012}
}

@article{Pozsgay_2013,
  title={The generalized {G}ibbs ensemble for {H}eisenberg spin chains},
  volume={2013},
  ISSN={1742-5468},
  url={http://dx.doi.org/10.1088/1742-5468/2013/07/P07003},
  DOI={10.1088/1742-5468/2013/07/p07003},
  number={07},
  journal={Journal of Statistical Mechanics: Theory and Experiment},
  publisher={IOP Publishing},
  author={Pozsgay, Balázs},
  year={2013},
  month=jul,
  pages={P07003}
}

@article{Prosen_2013,
  title={Families of Quasilocal Conservation Laws and Quantum Spin Transport},
  volume={111},
  ISSN={1079-7114},
  url={http://dx.doi.org/10.1103/PhysRevLett.111.057203},
  DOI={10.1103/physrevlett.111.057203},
  number={5},
  journal={Physical Review Letters},
  publisher={American Physical Society (APS)},
  author={Prosen, Tomaž and Ilievski, Enej},
  year={2013},
  month=aug
}

@article{Prosen_2014,
   title={Quasilocal conservation laws in {XXZ} spin-{$1/2$} chains: Open, periodic and twisted boundary conditions},
   volume={886},
   ISSN={0550-3213},
   url={http://dx.doi.org/10.1016/j.nuclphysb.2014.07.024},
   DOI={10.1016/j.nuclphysb.2014.07.024},
   journal={Nuclear Physics B},
   publisher={Elsevier BV},
   author={Prosen, Tomaž},
   year={2014},
   month=sep, pages={1177–1198}
}

@article{Richelli_2024,
  title={Brick wall quantum circuits with global fermionic symmetry},
  volume={17},
  ISSN={2542-4653},
  url={http://dx.doi.org/10.21468/SciPostPhys.17.3.087},
  DOI={10.21468/scipostphys.17.3.087},
  number={3},
  journal={SciPost Physics},
  publisher={Stichting SciPost},
  author={Richelli, Pietro and Schoutens, Kareljan and Zorzato, Alberto},
  year={2024},
  month=sep
}

@article{Sarma_2015,
  title={{M}ajorana zero modes and topological quantum computation},
  volume={1},
  ISSN={2056-6387},
  url={http://dx.doi.org/10.1038/npjqi.2015.1},
  DOI={10.1038/npjqi.2015.1},
  number={1},
  journal={npj Quantum Information},
  publisher={Springer Science and Business Media LLC},
  author={Sarma, Sankar Das and Freedman, Michael and Nayak, Chetan},
  year={2015},
  month=oct
}

@misc{Slavnov_2019,
  title={Algebraic {B}ethe ansatz},
  author={N. A. Slavnov},
  year={2019},
  eprint={1804.07350},
  archivePrefix={arXiv},
  primaryClass={math-ph},
  url={https://arxiv.org/abs/1804.07350},
}

@article{Sutherland_1970,
  author={Sutherland, Bill},
  title={Two‐Dimensional Hydrogen Bonded Crystals without the Ice Rule},
  journal={Journal of Mathematical Physics},
  volume={11},
  number={11},
  pages={3183-3186},
  year={1970},
  month={11},
  issn={0022-2488},
  doi={10.1063/1.1665111},
  url={https://doi.org/10.1063/1.1665111}
}

@article{Takhtadzhyan_1979,
  author={Takhtadzhyan, L. A. and Fadeev, L. D.},
  title={The quantum method of the inverse problem and the {H}eisenberg {XYZ} model},
  journal={Russian Math. Surveys},
  year=1979,
  volume=34,
  issue=5,
  pages={11--68},
  doi={10.1070/RM1979v034n05ABEH003909},
  url={http://mi.mathnet.ru/eng/rm4115}
}

@article{Teretenkov_2024,
  title={Exact dynamics of quantum dissipative {XX} models: {W}annier-{S}tark localization in the fragmented operator space},
  volume={109},
  ISSN={2469-9969},
  url={http://dx.doi.org/10.1103/PhysRevB.109.L140302},
  DOI={10.1103/physrevb.109.l140302},
  number={14},
  journal={Physical Review B},
  publisher={American Physical Society (APS)},
  author={Teretenkov, Alexander and Lychkovskiy, Oleg},
  year={2024},
  month=apr
}

@misc{Vanicat_2017,
  title={Integrable {T}rotterization: Local Conservation Laws and Boundary Driving},
  author={Matthieu Vanicat and Lenart Zadnik and Tomaž Prosen},
  year={2017},
  eprint={1712.00431},
  archivePrefix={arXiv},
  primaryClass={cond-mat.stat-mech},
  doi={10.1103/PhysRevLett.121.030606},
  url={https://arxiv.org/abs/1712.00431},
}

@misc{Vasseur_2026,
  title={Les {H}ouches lectures on random quantum circuits and monitored quantum dynamics},
  author={Romain Vasseur},
  year={2026},
  eprint={2602.17258},
  archivePrefix={arXiv},
  primaryClass={quant-ph},
  url={https://arxiv.org/abs/2602.17258},
}

@article{Vernier_2024,
  title={Strong Zero Modes in Integrable Quantum Circuits},
  volume={133},
  ISSN={1079-7114},
  url={http://dx.doi.org/10.1103/PhysRevLett.133.050606},
  DOI={10.1103/physrevlett.133.050606},
  number={5},
  journal={Physical Review Letters},
  publisher={American Physical Society (APS)},
  author={Vernier, Eric and Yeh, Hsiu-Chung and Piroli, Lorenzo and Mitra, Aditi},
  year={2024},
  month=aug
}

@article{Verstraete_2004,
  title={Matrix Product Density Operators: Simulation of Finite-Temperature and Dissipative Systems},
  volume={93},
  ISSN={1079-7114},
  url={http://dx.doi.org/10.1103/PhysRevLett.93.207204},
  DOI={10.1103/physrevlett.93.207204},
  number={20},
  journal={Physical Review Letters},
  publisher={American Physical Society (APS)},
  author={Verstraete, F. and García-Ripoll, J. J. and Cirac, J. I.},
  year={2004},
  month=nov
}

@article{Vidmar_2016,
  title={Generalized {G}ibbs ensemble in integrable lattice models},
  volume={2016},
  ISSN={1742-5468},
  url={http://dx.doi.org/10.1088/1742-5468/2016/06/064007},
  DOI={10.1088/1742-5468/2016/06/064007},
  number={6},
  journal={Journal of Statistical Mechanics: Theory and Experiment},
  publisher={IOP Publishing},
  author={Vidmar, Lev and Rigol, Marcos},
  year={2016},
  month=jun,
  pages={064007}
}

@article{Yang_1966_1,
  title={One-Dimensional Chain of Anisotropic Spin-Spin Interactions. {I.} Proof of {B}ethe's Hypothesis for Ground State in a Finite System},
  author={Yang, C. N. and Yang, C. P.},
  journal={Phys. Rev.},
  volume={150},
  issue={1},
  pages={321--327},
  numpages={0},
  year={1966},
  month={Oct},
  publisher={American Physical Society},
  doi={10.1103/PhysRev.150.321},
  url={https://link.aps.org/doi/10.1103/PhysRev.150.321}
}

@article{Yang_1966_2,
  title={One-Dimensional Chain of Anisotropic Spin-Spin Interactions. {II.} Properties of the Ground-State Energy Per Lattice Site for an Infinite System},
  author={Yang, C. N. and Yang, C. P.},
  journal={Phys. Rev.},
  volume={150},
  issue={1},
  pages={327--339},
  numpages={0},
  year={1966},
  month={Oct},
  publisher={American Physical Society},
  doi={10.1103/PhysRev.150.327},
  url={https://link.aps.org/doi/10.1103/PhysRev.150.327}
}

@article{Yamada_2023,
  title={Matrix product operator representations for the local conserved quantities of the {H}eisenberg chain},
  volume={6},
  ISSN={2666-9366},
  url={http://dx.doi.org/10.21468/SciPostPhysCore.6.4.069},
  DOI={10.21468/scipostphyscore.6.4.069},
  number={4},
  journal={SciPost Physics Core},
  publisher={Stichting SciPost},
  author={Yamada, Kyoichi and Fukai, Kouhei},
  year={2023},
  month=oct
}

@article{Yang_1966_3,
  title={One-Dimensional Chain of Anisotropic Spin-Spin Interactions. {III.} Applications},
  author={Yang, C. N. and Yang, C. P.},
  journal={Phys. Rev.},
  volume={151},
  issue={1},
  pages={258--264},
  numpages={0},
  year={1966},
  month={Nov},
  publisher={American Physical Society},
  doi={10.1103/PhysRev.151.258},
  url={https://link.aps.org/doi/10.1103/PhysRev.151.258}
}

@article{Yashin_2023,
  title={Integrable {F}loquet systems related to logarithmic conformal field theory},
  volume={14},
  ISSN={2542-4653},
  url={http://dx.doi.org/10.21468/SciPostPhys.14.4.084},
  DOI={10.21468/scipostphys.14.4.084},
  number={4},
  journal={SciPost Physics},
  publisher={Stichting SciPost},
  author={Yashin, Vsevolod I. and Kurlov, Denis V. and Fedorov, Aleksey K. and Gritsev, Vladimir},
  year={2023},
  month=apr
}

@article{Zadnik_2016,
  title={Quasilocal conservation laws from semicyclic irreducible representations of {$U_q(\mathfrak{sl}_2)$} in {XXZ} spin-{$1/2$} chains},
  volume={902},
  ISSN={0550-3213},
  url={http://dx.doi.org/10.1016/j.nuclphysb.2015.11.023},
  DOI={10.1016/j.nuclphysb.2015.11.023},
  journal={Nuclear Physics B},
  publisher={Elsevier BV},
  author={Zadnik, Lenart and Medenjak, Marko and Prosen, Tomaž},
  year={2016},
  month={jan},
  pages={339–353}
}

@article{Zadnik_2024,
  title={Quantum Many-Body Spin Ratchets},
  volume={5},
  ISSN={2691-3399},
  url={http://dx.doi.org/10.1103/PRXQuantum.5.030356},
  DOI={10.1103/prxquantum.5.030356},
  number={3},
  journal={PRX Quantum},
  publisher={American Physical Society (APS)},
  author={Zadnik, Lenart and Ljubotina, Marko and Krajnik, Ziga and Ilievski, Enej and Prosen, Tomaž},
  year={2024},
  month=sep
}

@article{Znidaric_2025,
  title={Integrability is generic in homogeneous {$U(1)$}-invariant nearest-neighbor qubit circuits},
  volume={112},
  ISSN={2469-9969},
  url={http://dx.doi.org/10.1103/tqy8-ynpd},
  DOI={10.1103/tqy8-ynpd},
  number={2},
  journal={Physical Review B},
  publisher={American Physical Society (APS)},
  author={Žnidarič, Marko and Duh, Urban and Zadnik, Lenart},
  year={2025},
  month=jul
}

@article{Zwolak_2004,
  title={Mixed-State Dynamics in One-Dimensional Quantum Lattice Systems: A Time-Dependent Superoperator Renormalization Algorithm},
  volume={93},
  ISSN={1079-7114},
  url={http://dx.doi.org/10.1103/PhysRevLett.93.207205},
  DOI={10.1103/physrevlett.93.207205},
  number={20},
  journal={Physical Review Letters},
  publisher={American Physical Society (APS)},
  author={Zwolak, Michael and Vidal, Guifré},
  year={2004},
  month=nov
}

@misc{Mathematica,
  author={{Wolfram Research{,} Inc.}},
  title={Mathematica, {V}ersion 14.3},
  url={https://www.wolfram.com/mathematica},
  note={Champaign, IL, 2025}
}

@dataset{Supplementary,
  author={Yashin, V. I.},
  title={Supplementary Materials for {V.I. Yashin} ``{T}wo-parameter families of {MPO} integrals of motion in {H}eisenberg spin chains''},
  month=mar,
  year=2026,
  publisher={Zenodo},
  doi={10.5281/zenodo.19212085},
  url={https://doi.org/10.5281/zenodo.19212085}
}

@misc{Private,
  author={Fukai, Kohei},
  title={Private communication.}
}


\end{document}